\documentclass{birkjour_t2}

\usepackage[latin2]{inputenc}
\usepackage{amsmath}
\usepackage{amssymb}
\usepackage{amsfonts}
\usepackage{stmaryrd}
\usepackage{graphicx}
\usepackage[parfill]{parskip}    
\usepackage{hyperref}
\hypersetup{pdftex,colorlinks=true,allcolors=blue}

\newtheorem{thm}{Theorem}[section]
\newtheorem{lem}{Lemma}[section]

\newtheorem{cnj}{Conjecture}[section]

\newtheorem{dfn}{Definition}[section]

\newtheorem{prb}{Problem}[section]

\begin{document}


\title[Fractal Potential Flows]{Fractal Potential Flows\\{\Large An Idealized Model for Fully Developed Turbulence}}
\author{József Vass}
\address{Department of Applied Mathematics\\
        University of Waterloo\\
        200 University Avenue West\\
        Waterloo, ON, N2L 3G1, Canada}
\email{jvass@uwaterloo.ca}
\date{\displaydate{date}}

\newcommand*{\me}{\mathrm{e}}
\newcommand*{\vi}{\mathrm{i}}
\newcommand*{\mmod}{\mathrm{mod}}
\newcommand*{\argmin}{\mathrm{argmin}}
\newcommand*{\argmax}{\mathrm{argmax}}
\newcommand*{\Argmax}{\mathrm{Argmax}}
\newcommand*{\adr}{\mathrm{adr}}
\newcommand*{\diam}{\mathrm{diam}}
\newcommand*{\Conv}{\mathrm{Conv}}
\newcommand*{\Span}{\mathrm{Span}}
\newcommand*{\Nei}{\mathrm{Nei}}
\newcommand*{\Ext}{\mathrm{Ext}}
\newcommand*{\Per}{\mathrm{Per}}
\newcommand*{\Eve}{\mathrm{Eve}}
\newcommand*{\Foc}{\mathrm{Foc}}
\newcommand*{\Cyc}{\mathrm{Cyc}}
\newcommand*{\Cl}{\mathrm{Cl}}
\newcommand*{\msgn}{\mathrm{sgn}}
\newcommand*{\mint}{\mathrm{int}}
\newcommand*{\msupp}{\mathrm{Supp}}
\newcommand*{\Ran}{\mathrm{Ran}}
\newcommand*{\Ker}{\mathrm{Ker}}
\newcommand*{\mRe}{\mathrm{Re}}
\newcommand*{\mIm}{\mathrm{Im}}
\newcommand*{\mdiv}{\mathrm{div}}
\newcommand*{\mcurl}{\mathrm{curl}}
\newcommand*{\mleb}{\mathrm{leb}}
\newcommand*{\mae}{\mathrm{a.e.}}
\newcommand*{\Arg}{\mathrm{Arg}}
\newcommand*{\Log}{\mathrm{Log}}
\newcommand*{\mand}{\mathrm{and}}
\newcommand*{\tr}{\mathrm{tr}}
\newcommand*{\trn}{\mathrm{trn}}
\newcommand*{\chr}{\mathrm{char}}
\newcommand*{\T}{\mathrm{T}}
\newcommand*{\D}{\mathrm{D}}
\newcommand*{\md}{\mathrm{d}}
\newcommand*{\mP}{\mathrm{Pr}}
\newcommand*{\mE}{\mathrm{E}}
\newcommand*{\mH}{\mathrm{H}}
\newcommand*{\mtrue}{\mathrm{true}}
\newcommand*{\mfalse}{\mathrm{false}}
\newcommand*{\cf}{\ensuremath{\varphi}}
\newcommand*{\vr}{\ensuremath{\varrho}}
\newcommand*{\ve}{\ensuremath{\varepsilon}}
\newcommand*{\vt}{\ensuremath{\vartheta}}
\newcommand*{\vd}{\ensuremath{\delta}}
\newcommand*{\vl}{\ensuremath{\lambda}}
\newcommand*{\va}{\ensuremath{\alpha}}
\newcommand*{\vs}{\ensuremath{\sigma}}
\newcommand*{\vm}{\ensuremath{\mu}}
\newcommand*{\grd}{\ensuremath{\nabla}}
\newcommand*{\lap}{\ensuremath{\triangle}}
\newcommand*{\nub}{\ensuremath{\bar{\nu}}}
\newcommand*{\psib}{\ensuremath{\bar{\psi}}}
\newcommand*{\N}{\ensuremath{\mathbb{N}}}
\newcommand*{\Z}{\ensuremath{\mathbb{Z}}}
\newcommand*{\Q}{\ensuremath{\mathbb{Q}}}
\newcommand*{\R}{\ensuremath{\mathbb{R}}}
\newcommand*{\C}{\ensuremath{\mathbb{C}}}
\newcommand*{\prf}{\textbf{Proof}\ \ }
\newcommand*{\sln}{\textbf{Solution}\ \ }
\newcommand*{\rea}{\textbf{Reasoning}\ \ }
\newcommand*{\sqr}{\ensuremath{\square}}
\newcommand*{\rar}{\ensuremath{\Rightarrow}}
\newcommand*{\frar}{\ensuremath{\shortrightarrow}}
\newcommand*{\lar}{\ensuremath{\Leftarrow}}
\newcommand*{\lrar}{\ensuremath{\Leftrightarrow}}
\newcommand*{\mcN}{\ensuremath{\mathcal{N}}}
\newcommand*{\mcT}{\ensuremath{\mathcal{T}}}
\newcommand*{\mcP}{\ensuremath{\mathcal{P}}}
\newcommand*{\mcA}{\ensuremath{\mathcal{A}}}
\newcommand*{\mcAf}{\ensuremath{\mathcal{A}_{fin}}}
\newcommand*{\mcAi}{\ensuremath{\mathcal{A}_\infty}}
\newcommand*{\MX}{\ensuremath{\mathcal{M}_X}}
\newcommand*{\PX}{\ensuremath{\mathcal{P}_X}}
\newcommand*{\DX}{\ensuremath{\mathcal{D}_X}}
\newcommand*{\Lip}{\ensuremath{\mathcal{L}_1(X)}}
\newcommand*{\BCci}{\ensuremath{BC_c^\infty}}
\newcommand*{\PsXc}{\ensuremath{\Psi_{X,c}}}
\newcommand*{\bPsXc}{\ensuremath{[\Psi_{X,c}]}}
\newcommand*{\qc}{\ensuremath{q_c}}
\newcommand*{\gc}{\ensuremath{\Gamma_c}}
\newcommand*{\co}{\ensuremath{\circ}}
\newcommand*{\bu}{\ensuremath{\bullet}}
\newcommand*{\tti}{\ensuremath{\rightarrow\infty}}
\newcommand*{\mes}{\ensuremath{\varnothing}} 
\newcommand*{\eqdef}{\mathrel{\overset{\makebox[0pt]{\mbox{\normalfont\tiny\sffamily def}}}{=}}}
\newcommand{\mpbar}{\smash{\rule[-1.4ex]{0.4pt}{4ex}}} 

\begin{abstract}
Fully Developed Turbulence (FDT) is a theoretical asymptotic phenomenon which can only be approximated experimentally or computationally, so its defining characteristics are hypothetical. It is considered to be a chaotic stationary flow field, with self-similar fractalline features. A number of approximate models exist, often exploiting this self-similarity. The idealized mathematical model of Fractal Potential Flows is hereby presented, and linked philosophically to the phenomenon of FDT on a free surface, based on its experimental characteristics. The model hinges on the recursive iteration of a fluid dynamical transfer operator. The existence of its unique attractor - called the invariant flow - is shown in an appropriate function space, which will serve as our suggested model for the FDT flow field. Its sink singularities are shown to form an IFS fractal, explicitly resolving Mandelbrot's Conjecture. Meanwhile an isometric isomorphism is defined between flows and probability measures, hinting at a wealth of future research. The inverse problem of representing turbulent flow fields with this model is discussed in closing, along with explicit practical considerations for experimental verification and visualization.
\end{abstract}

\thanks{The model was presented at the 2012 Winter Meeting of the Canadian Mathematical Society. The first draft of the paper was shared in December 2012, while the second draft was finalized in September 2014.}

\maketitle



\section{Overview} \label{s01}

An overview of Fully Developed Turbulence is presented in Section \ref{s0201}, in an attempt to identify the definitive characteristics of this hypothetical phenomenon at near-infinite Reynolds numbers, which can only be approximated with experiments or computer simulation. We proceed in Section \ref{s0203} to the unconventional direction of non-deterministic experimental Fluid Dynamics called Chaotic Advection, to be reformulated in the language of Optimal Transport Theory. Section \ref{s0202} gives an introduction to Potential Flows and their superposition, intriguingly sharing its inviscid and stationary characteristics with FDT. Since the well-known IFS invariant measure is the fixed point of the iterative evolution under the measure transfer operator, it will serve as an inspiration for showing the unique existence of an invariant flow, using the concepts of Sections \ref{s0204} and \ref{s0205}. Since the transfer operator distributes singularities over an IFS fractal, this adds impetus to pursuing a transfer model, considering Mandelbrot's Conjecture. In Section \ref{s0206}, a decisive step is made to abandon Navier-Stokes evolution on mathematical grounds, observing the non-smooth nature of intermittency, so we focus on discrete inertial transfer evolution of piecewise steady flows in the spirit of Chaotic Advection / Mixing.

Building on the reviewed theories, the idealized model of Fractal Potential Flows is presented in Section \ref{s03}. The random intermittent interaction of a finite system of eddies is directly modelled via the recursive iteration of a weighted transfer operator in Section \ref{s0302}, inspired by the invariance of an eddy in Section \ref{s0301}. The contractive action of each eddy sink on the flow field, is modelled with a corresponding pushforward transfer map, induced by an affine contraction map of the plane. Since the collective action of the eddies is probabilistically weighted, it can be interpreted as an ``expected value'' expression, while physically speaking as the weighted effect each eddy has on the flow field. The iteration of the transfer operator is reasoned to correspond to the non-smooth discrete energy bursts of intermittency.

The evolution equation towards FDT having been reasoned to be the discrete recursion of a transfer operator, our goal consequently becomes to show that it converges to a unique attractor flow - the FDT flow field itself - in some appropriate function space, revealed in Section \ref{s0304}. In finding the correct space, eddies remain our guiding inspiration, considering that a system of eddies becomes a superposed eddy when zoomed out, possessing a complex character at complex infinity as discussed in Section \ref{s0303}. In their local universe, the eddies interact to produce an increasingly fractalline flow field, approaching the desired invariant flow shown in Section \ref{s0305}. Lastly, the sinks of the invariant flow field are shown to form an IFS fractal in Section \ref{s030601}, implying the ultimate conclusion that the geometrical study of such fractals, is of fundamental relevance to analyzing an FDT flow field.

An adaptation of the inverse problem of IFS to fractal potential flows is described in Section \ref{s0403}. The elements of the model that require experimental verification are isolated in Section \ref{s0404}. Lastly, our method of visualization is detailed in Section \ref{s0401}.
\section{Preliminary Concepts} \label{s02}

\subsection{Fully Developed Turbulence} \label{s0201}

\subsubsection{Characteristics, Causes, and Evolution} \label{s020101}

Turbulence has been the interest of classical physics for centuries. Its name originates in the Italian word ``turbolenza'' given by Leonardo da Vinci, one of the first to study and depict the phenomenon \cite{bb00017}. Turbulence is a challenging phenomenon to model, primarily because it is difficult to define rigorously. One must isolate from experimental observation its definitive phenomenological characteristics, prior to attempting to model it. A rich literature exists on the topic of FDT, and the overview of Ecke \cite{ba00032} is recommended for an introduction, while those of Frisch \cite{ba00087, bb00016} for more details.

In terms of the Reynolds number \cite{ba00085}, flows above an $Re$ of $5000$ or higher are typically considered to be turbulent. The phenomenon of Fully Developed Turbulence is an asymptotic case of turbulence, based on experimental observations at high Reynolds numbers, typically above $10,000$. The words ``fully developed'' in FDT refer to its main theoretical characteristics: a near-infinite Reynolds number and a stationary (statistically unchanging) yet chaotic velocity field, in which viscous effects become negligible (the Navier-Stokes Equations thus reduce to the Euler Equations) \cite{bp00004, bp00005, bp00006}. This signifies an equilibrium state, or invariance with respect to the evolution process (in the model to be presented, the iteration of an operator).

Further characteristics include high and erratic velocity fluctuation over a large range of coupled spatial and temporal scales \cite{ba00032}. Consequently, a turbulent flow field exhibits structure at many length scales, often described as self-similar, fractalline, or scale-independent in some sense \cite{bb00015}. According to Orszag et al. its chaos can be spatiotemporal \cite{ba00088, bp00007}. Frisch et al. \cite{ba00077} specifically note that intermittency is necessarily also temporal because of the sweeping of large structures by smaller ones, but there may exist an intrinsic temporal intermittency as well \cite{ba00081, ba00095, ba00096}. Frisch \cite{ba00087} remarks that the evolutionary dynamical system can be chaotic while the velocity field remains steady, and only the streamlines reflect chaos - an example being the Arnold-Beltrami-Childress Flow. This type of ``Lagrangian turbulence'' can manifest in two dimensions, as shown by Aref \cite{ba00034, ba00035} (also see Section \ref{s030602}).

The onset and evolution of turbulence can be fuelled by various circumstances, such as the interaction of eddies in the flow, by injected kinetic energy into the flow such as via stirring, or even by the shape or surface of the container. Examples include flow from a faucet, flow past rocks, or the rotation of the Earth \cite{ba00032}. A periodicity in the circumstances often plays a role in inducing ``intermittency''. Mathematically speaking, development can occur due to periodic changes in the external forcing, or variations in the boundary conditions. In intermittency, transitions are observed to occur abruptly to successively more complex states, discovered by Batchelor and Townsend \cite{ba00092}. The evolution of turbulence is thought to be a purely inertial process, meaning it is due to the actual motion of the fluid, while flows where inertial effects are small, tend to remain laminar (non-turbulent).

Eddies and their intermittent inertial interaction are considered to be fundamental elements of turbulent dynamics - possibly the root cause of the evolution to FDT. Since the injected energy at the top scale is periodic, the eddies in their evolutionary hierarchy receive periodic bursts of energy as well (this is the Richardson energy cascade \cite{ba00094, bb00015}), causing their interaction to be intermittent. In between each energy burst, a brief quiescent period occurs, where the fluid flows along its natural course i.e. its steady streamlines \cite{ba00032, ba00092}.

A hypothetical picture of the eddy hierarchy, is the interplay of vortex stretching detailed by Taylor \cite{ba00093} and the energy cascade theorized by Richardson \cite{ba00094, bb00015}. Vortex stretching occurs due to the conservation of fluid elements, and this results in the breakdown of larger structures into hierarchically smaller ones, causing an intermittent cascade of energy \cite{ba00081}. This is a purely inertial process, which continues until the local structures are small enough that viscosity causes their kinetic energy to dissipate as heat. Note that according to experimental evidence, the eddies of FDT become less and less space-filling as the scale decreases \cite{ba00077, ba00092, ba00097, bb00018}.

\begin{cnj} \label{s02010301} \textup{(Mandelbrot \cite{bc00008})}
At infinite Reynolds numbers (FDT), the energy dissipation of the fluid concentrates in a set of non-integer Hausdorff dimension.
\end{cnj}

This conjecture has been resolved by Scheffer in 1980 \cite{ba00086}, though the proof is non-constructive. If one wishes to construct an FDT flow field, this conjecture may serve as an inspiration.

\subsubsection{Existing Models} \label{s020102}

Former efforts to model turbulence in general, and FDT in particular, have been approximate or statistical in nature, mostly in the form of scaling laws derived from the Navier-Stokes Equations. This direction was inspired by Kolmogorov's influential model based on the Richardson cascade \cite{bp00004, bp00005, bp00006, ba00082}, culminating in the current Beta Model developed by Novikov and Stewart \cite{ba00079}, Kraichan \cite{ic00007, ba00081}, Mandelbrot \cite{ba00080, ic00006}, Frisch et al. \cite{ba00077, bb00016}, and many others (for a brief survey, see Jou \cite{ba00033}). Further significant study has been done by McComb \cite{ba00083} and Sreenivasan \cite{ba00084}. The computational direction of simulating large Reynolds number flows on high-performance computers (for Large Eddy Simulation, see Meneveau and Katz \cite{ba00078}) was originally suggested by John von Neumann \cite{ic00005}. Ruelle \cite{ba00090} takes a measure theoretic approach to the study of turbulence, which is close in sentiment to our own, as is his dynamical systems approach with Takens \cite{ba00091}.

The theory of Chaotic Advection is an alternative experimental and mathematical direction. It is a dynamical systems approach to turbulence, which investigates the intermittent interaction of eddies directly. See the next section for an overview.

Rigorous understanding and prediction from first principles through an experimentally verified model is yet to be achieved for the phenomenon of Fully Developed Turbulence. Nevertheless, a new model must only be in experimental agreement with the phenomenon itself, and not with other reasoned models. In other words, we only need to consider the phenomenon as observed in experiments when attempting to build a new model, which must then be contrasted with the phenomenon via further experiments.
\newpage
\subsection{Chaotic Advection} \label{s0203}

\subsubsection{Aref's Blinking Vortex-Sink System} \label{s020301}

Advection is the idea of a fluid transporting light matter - typically tracer particles - on its free surface, so that the velocity of the particles is given by the velocity field of the flow. When $\psi$ is the stream function of a velocity field $v$, the Hamiltonian equations $v_1 = -\partial_2\psi,\ v_2 = \partial_1\psi$ hold. Hence the configuration space of an advected particle is the phase space of this Hamiltonian system. Often such a system exhibits chaos, or sensitive dependence on initial conditions. The notion that laminar flows can produce chaotic particle trajectories was pioneered and advocated by Hassan Aref \cite{ba00034, ba00035}. The theory essentially reproduces by stirring the effect of on-off intermittency discovered experimentally by Batchelor and Townsend \cite{ba00092}. See Ottino \cite{bb00008} for an overview of Chaotic Advection / Mixing, and \cite{ba00076} for reasoned connections to turbulence.

The flows studied in this context are often piecewise steady. Meaning that on the evolution timeline of the flow, the velocity field is steady for consecutive time intervals, and a non-smooth abrupt transition (jump) to a new velocity field occurs at each partitioning moment in time. The jumps act along the streamlines of two circular (or logarithmic) vortices, which take alternating (non-random) turns in perturbing the flow field. This is called the Blinking Vortex(-Sink) System, and has been examined extensively by a number of theorists and experimentalists. It is meant to directly simulate the intermittent interaction of (circular or) logarithmic vortices, which is considered to be the elementary driving mechanism of the evolution towards Fully Developed Turbulence. Aref in his highly influential paper \cite{ba00034} originally defined the setup with two circular vortices, which has been modified in \cite{ba00035} to two vortex-sinks (eddies), and was further studied by Károlyi and Tél \cite{ba00036} as well as a number of other researchers.

In particular, Wiggins \cite{ba00043} discusses the method and implications of modeling this blinking as the alternating (non-random) infinite iterative composition of two maps (called linked twist maps, the study of which goes back to Devaney \cite{bp00001}) in order to show the existence of a chaotic invariant set via the Conley-Moser conditions. These ideas are further examined with Ottino in \cite{ba00044}, and bear an inspirational resemblance to our own, to be presented in this paper.

\subsubsection{Ott's Chaos Game} \label{s020302}

Ott et al. \cite{ba00038} made a significant improvement on Aref's model. The regular iterative alternation between two vortex-sinks has been modified to a random iteration over a finite family of mappings that transform the flow at intermittent time partitions, in between steady flow periods. So at step $l\in\N$ (time partition $t_l$) the state of the flow $\sigma_l$ (usually the position of a tracer particle on the fluid surface) is transformed as $\sigma_{l+1} = T_l(\sigma_l)$ where $T_l\in\{T_1,\ldots,T_n\}$ is chosen at random, and where the precise form of $T_l$ can vary and is often left implicit. This random discrete evolution is reminiscent of the Chaos Game of Barnsley \cite{bb00002}. Though this formulation can be closely aligned with experiments, it is difficult however to treat it mathematically. Ott et al. further assume that $t_l$ are spaced evenly and their stream functions are time-dependent, as typical in the theory of Chaotic Advection. However as we will see, both of these are unnecessary constraints, in our explicit formulation of this chaos game in Section \ref{s0302}. An overview of the work of Ott and Sommerer et al. in collaboration with Tél has been presented in \cite{ba00037}, building upon \cite{ba00038, ba00039, ba00040}.

\subsubsection{Sommerer's Experiment} \label{s020303}

Chaotic Advection is a de facto experimental theory, and its predictions have been verified by both computational and real-world experiments. Sommerer's apparatus \cite{ba00039, ba00037} carried out at the JHU Applied Physics Laboratory, consists of a tank of water with a set of sources at the bottom, which transmit pulsing injections from an external pulsatile flow modulator that randomly regulates which source receives the next injection at equal time intervals, circulated by an external pump. Each pulse corresponds to an iterative step $\sigma_{l+1} = T_l(\sigma_l)$ where the fixed set of $T_l\in\{T_1,\ldots,T_n\}$ represents the actions of each source on the particles. These fluorescent particles on the fluid surface are then traced over their chaotic paths. In a sense, this apparatus is an inversion of the Aref setup, since it utilizes sources instead of sinks. Nevertheless, its relevance lies in attempting to simulate the intermittent evolution of turbulence, in the spirit of Ott's chaos game recursion.

\subsection{Potential Flows} \label{s0202}

For flows with low viscosity outside the boundary layer, potential flows are considered to provide a sufficient model. Yet the turbulent boundary layer itself is often modelled approximately with a superposition of vortices (Vortex Dynamics). For FDT flow fields viscosity becomes negligible, so potential flows may play a role in their modeling. Potential flow is assumed to be steady (time-independent), ideal (zero viscosity / inviscid, uniform density, and incompressible), and irrotational (almost everywhere with respect to the Lebesgue measure $\mleb$, denoted as a.e.). We derive the stream and potential functions of the superposition of a sink and a vortex resulting in an eddy \cite{bb00013, bb00014}.

By the above conditions, conservation of mass for a velocity field $v=(v_1,v_2):\R^2 \frar \R^2$ requires
\[ 0 = \mdiv\ v = \partial_1 v_1 + \partial_2 v_2\ \ \mae \]
and the equation of motion becomes
\[ \frac12\grd |v|^2 - v\times\mcurl\ v = -\frac{1}{\vr}\grd p \]
where $\vr$ is the density and $p$ is pressure, further reducing to
\[ p(z) = \left(p(z_0)+\frac{\vr}{2} |v(z_0)|^2\right) -  \frac{\vr}{2} |v(z)|^2 \]
under the irrotationality requirement
\[ 0 = \mcurl\ v = \partial_2 v_1 - \partial_1 v_2\ \ \mae \]

Supposing that $\psi,\phi\in C^2_{ae}(\R^2,\R)$ are harmonic conjugates, meaning they satisfy the Laplace and Cauchy-Riemann Equations
\[ \lap\psi=\lap\phi=0\ \mae\ \ \ \mathrm{and}\ \ \ \partial_1\phi = \partial_2\psi,\ \partial_2\phi = -\partial_1\psi\ \mae \]
then $v:= -\grd\phi = \vi\grd\psi$ satisfies mass conservation and irrotationality $\mae$ The pressure field can be calculated from $v$ as above. We call such an $\mae$ harmonic $\psi$ for which a conjugate exists, the stream function of a potential flow, and $\phi$ the potential function. Note that the potential function corresponding to a stream function is only unique up to a gradient. Nevertheless we denote it as $\tilde{\psi}:=\phi$, and denote equivalence in the gradient $\mae$ as
\[ \phi_1\equiv\phi_2\ \lrar\ \grd\phi_1=\grd\phi_2\ \mae \]
A harmonic function is known to admit a conjugate if its domain is simply connected. Furthermore $\tilde{\tilde{\psi}}=-\psi$. By their relation to the velocity field, the curves of constant $\psi$ represent the streamlines of the flow, and the curves of constant $\phi$ the equipotential lines. Having two pairs of harmonic conjugates $\psi_1, \phi_1$ and $\psi_2, \phi_2$ their linear combinations $a\psi_1 + b\psi_2$ and $a\phi_1 + b\phi_2$ for any $a,b\in\R$ are also harmonic conjugates. So the principle of superposition holds, as long as the boundary conditions are also correspondingly combined. Since the equations of motion have a unique solution for a set of boundary conditions, and they translate to Laplace's equation (the solution of which for appropriate boundary conditions also exists and is unique up to a gradient), we may conclude that potential flows fully characterize the set of all flows which are steady, ideal, and irrotational $\mae$

Elementary potential flows, which are often superposed to create more complex ones, include sinks, sources and the (circular) vortex. The stream and potential functions of a source ($q>0$) or a sink ($q<0$) with strength $q\in\R$ are defined at $z\in\C$ as
\[ \psi(z) = -\frac{q}{2\pi}\Arg\ z\ \mmod\ |q|\ \ \mathrm{and}\ \ \phi(z) = -\frac{q}{2\pi}\ln |z| \]
while for the orthogonal circular vortex
\[ \psi(z) = \frac{\Gamma}{2\pi}\ln |z|\ \ \mathrm{and}\ \ \phi(z) = -\frac{\Gamma}{2\pi}\Arg\ z\ \mmod\ |\Gamma| \]
where $\Gamma$ is the circulation around any closed path containing the vortex (strength of the vortex). Superposing a sink and a circular vortex with a general center $p\in\C$ results in a logarithmic vortex (eddy) as follows
\[ \psi(z) = -\frac{q}{2\pi}\Arg(z-p) + \frac{\Gamma}{2\pi}\ln |z-p|\ \ \mmod\ |q| \]
\[ \phi(z) = -\frac{q}{2\pi}\ln |z-p| - \frac{\Gamma}{2\pi}\Arg(z-p)\ \ \mmod\ |\Gamma|. \]
Since $\psi$ is differentiable in the $\R^2$ sense almost everywhere (except at $p$ and along the branch cut $\psi(z)=0$) we have that the corresponding velocity field (extended continuously to the entire plane) is the following
\[ v(z) = \frac{q+\Gamma\vi}{2\pi}\frac{1}{|z-p|}\frac{z-p}{|z-p|}\ \ (z\in\C). \]
Here the gradient vectors are considered to be on the complex plane, with differentiation in the bivariate sense, meaning $\grd\psi = \partial_1\psi +\vi\partial_2\psi$. In general, we sloppily identify $z\mapsto\psi(z),\ z\in\C$ with $(x,y)\mapsto\psi(x+y\vi),\ (x,y)\in\R^2$ when it is more convenient.

\subsection{IFS Fractals} \label{s0204}

\begin{figure}[H]
\centering
\includegraphics[width=410pt]{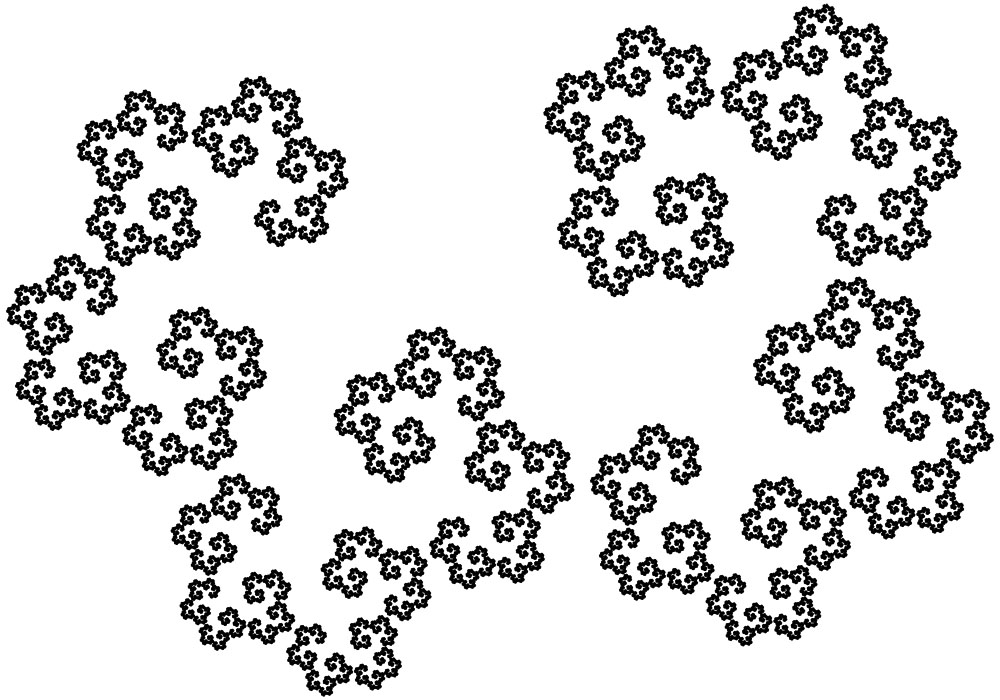}\\
\vspace{0.5cm}
\includegraphics[width=410pt]{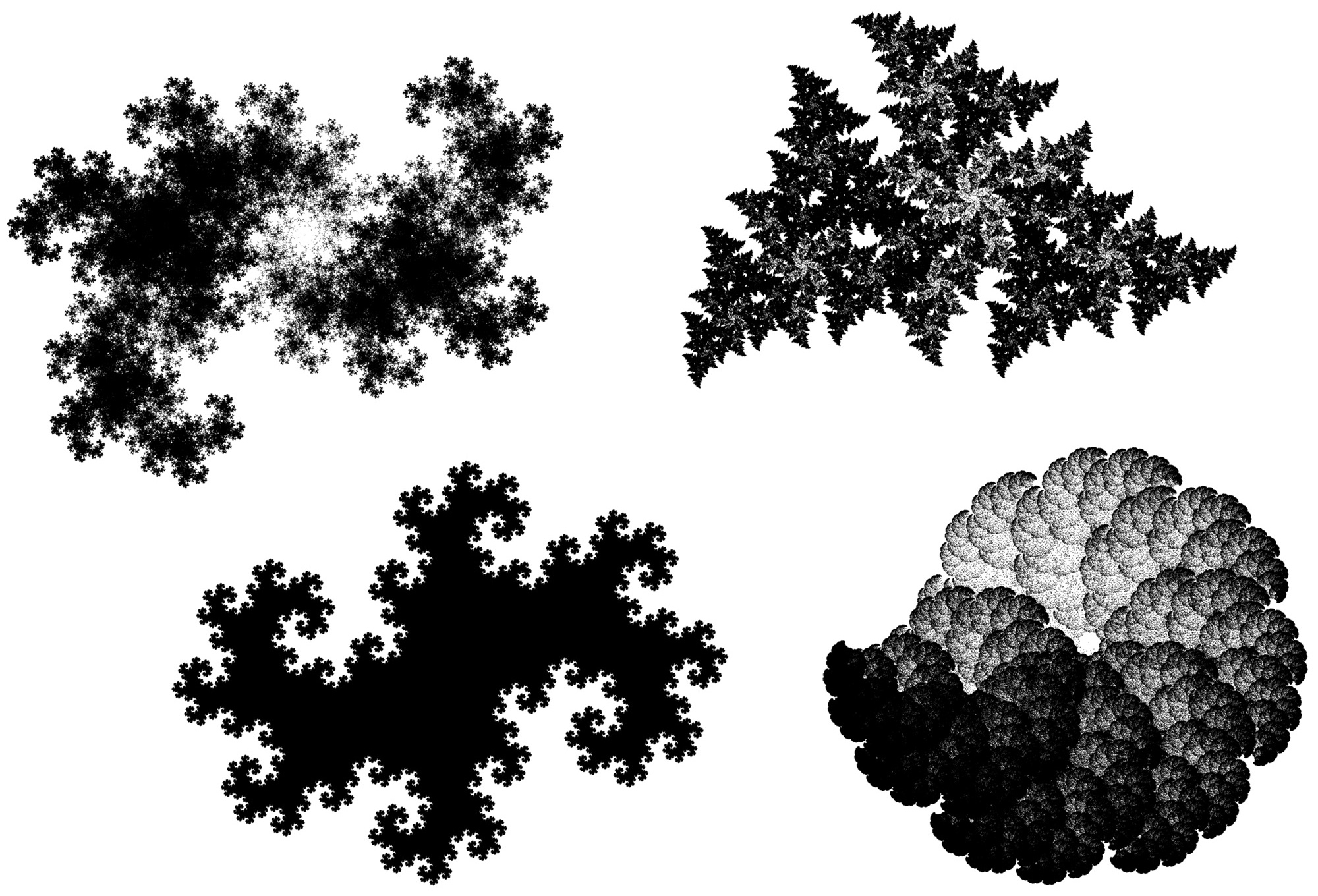}\\
\vspace{0.4cm}
\caption{The IFS fractal of sink singularities for the parameters in Section \ref{s0401}, and some more examples of such fractals, studied in \cite{ph00004}.}
\label{s04010005}
\end{figure}

The attractors of Iterated Function Systems - IFS fractals - were pioneered by Hutchinson \cite{ba00007}, further discussed by Barnsley and Demko \cite{ba00017} as well as others, and in a sense may be the most elementary fractals possible. They are the attractors of a finite set of affine linear contraction mappings on the plane - the Iterated Function System (IFS) - which when combined and iterated to infinity, converges to an attracting limit set, the IFS fractal itself (see Fig. \ref{s04010005} for some examples).

\subsubsection{Definition and Existence} \label{s020401}

\begin{dfn} \label{s02040101}
Let a planar similarity affine contractive mapping (briefly similarity contraction, contraction map, or similitude) $T:\C\shortrightarrow\C$ be defined for all $z\in\C$ as $T(z):=p+\varphi(z-p)$ where $p\in\C$ is the fixed point, $\cf=\vl e^{\vt i}\in\C$ is the factor, with $\vl\in(0,1)$ is the contraction factor, and $\vt\in(-\pi,\pi]$ is the rotation angle of $T$.
\end{dfn}

Note that an equivalent definition may be given using unitary rotation matrices $R\in\R^{d\times d},\ R^{T}\!R=I,\ d\in\N$ corresponding to $e^{\vt i}$ when $d=2$. Then contraction maps take the following form
\[ T(z)=p+\vl R(z-p)\ \ \ (z\in\R^d,\ p\in\R^d,\ \vl\in (0,1)). \]
This form shall be useful for certain proofs in later sections.

\begin{dfn} \label{s02040102}
Let a planar similarity affine contractive $n$-map iterated function system (briefly IFS or $n$-map IFS, $n\in\N$) be defined as a finite set of contractions, and denoted as $\mathcal{T}:=\{T_1,\ldots,T_n\}$. Denote the set of indices as $\mcN:=\{1,\ldots,n\}$, the respective fixed points as $\mcP:=\{p_1,\ldots,p_n\}$, and the factors as $\cf_1,\ldots,\cf_n$.
\end{dfn}

\begin{dfn} \label{s02040103}
Define the Hutchinson operator $\mH$ belonging to the IFS $\mcT=\{T_1,\ldots,T_n\}$ as
\[ \mH(S)=\mH_\mcT(S):=\bigcup_{k=1}^n T_k(S)\ \ \mathrm{where}\ \ T_k(S):=\{T_k(z): z\in S\}\ \ \mathrm{for\ any}\ \ S\subset\C \]
and call $\mH(S)$ the Hutchinson of the set $S$ with respect to $\mcT$.
\end{dfn}

\begin{thm} \label{s02040104} \textup{(Hutchinson \cite{ba00007})}
For any IFS $\mcT$ with Hutchinson operator $\mH_\mcT$, there exists a unique compact set $F_\mcT\subset\C$ such that $\mH_\mcT(F_\mcT)=F_\mcT$. Furthermore, for any nonempty compact set $S_0\subset\C$, the recursive iteration $S_{n+1}:=\mH_\mcT(S_n)$ converges to $F_\mcT$ in the Hausdorff metric.
\end{thm}\begin{proof}
The proof follows from the Banach Fixed Point Theorem, since it can be shown that $\mH_\mcT$ is contractive in the Hausdorff metric over the complete metric space of nonempty compact subsets.
\end{proof}

\begin{dfn} \label{s02040105}
Let the set $F_\mcT$ in the above theorem be called a fractal generated by an IFS $\mcT$ (briefly IFS fractal). Denote $\langle\mathcal{T}\rangle = \langle T_1,\ldots,T_n\rangle :=F_\mcT$.
\end{dfn}

\subsubsection{The Address Set} \label{s020402}

The address set arises from the iteration of the Hutchinson operator, and it is a way to label each fractal point. Since the iteration towards $F$ can be started with any compact set, typically the primary fixed point is chosen for simplicity, which is any point in $\mcP$ of our preference.

\begin{dfn} \label{s02040201}
Let $\mcN^L:=\mcN\times\ldots\times\mcN$ be the index set to the $L$-th Cartesian power, and call this $L\in\N$ the iteration level. Then define the address set as
\[ \mcA:=\{0\}\cup\bigcup_{L=1}^\infty\mcN^L\cup\mcN^\N. \]
For any $a\in\mcA$ denote its $k$-th coordinate as $a(k),\ k\in\N$. Let its dimension or length be denoted as $|a|\in\N$ so that $a\in\mcN^{|a|}$ and let $|0|:=0$. Define the map with address $a\in\mcA$ acting on any $z\in\C$ as the function composition $T_a(z):=T_{a(1)}\circ\ldots\circ T_{a(|a|)}(z)$. Let the identity map be $T_0:=\mathrm{Id}$. Denote
\[ \mcAf:=\{a\in\mcA: |a|<\infty\},\ \mcAi:=\{a\in\mcA: |a|=\infty\}=\mcN^\N. \]
For the weights $w_1,\ldots,w_n\in (0,1)$ let $w_a:= w_{a(1)}\cdot\ldots\cdot w_{a(|a|)}$, for the factors $\cf_1,\ldots,\cf_n\in\C$ let $\cf_a:= \cf_{a(1)}\cdot\ldots\cdot\cf_{a(|a|)}$, and for the angles $\vt_1,\ldots,\vt_n\in (-\pi,\pi]$ let $\vt_a:=\vt_{a(1)}+\ldots +\vt_{a(|a|)}$.
\end{dfn}

\begin{thm} \label{s02040202}
For any primary fixed point $p\in\mcP$ the fractal can be generated as
\[ \langle T_1,\ldots,\T_n\rangle = \lim_{L\rightarrow\infty} \mH^L(\{p\}) = \mathrm{Cl}\{T_a(p):a\in\mcAf\} = \mathrm{Cl}\{T_a(p_k):a\in\mcAf,\ p_k\in\mcP\}. \]
We call this the address generation of the IFS fractal $\langle T_1,\ldots,\T_n\rangle$.
\end{thm}\begin{proof}
The proof follows from Theorem \ref{s02040104} with the initial sets $\{p\}$ or $\mcP$.
\end{proof}

\begin{dfn} \label{s02040203}
Inspired by the above theorem, we will call $F_L:=\{T_a(p): a\in\mcA, |a|=L\}$ (for some $L\in\N$ and primary fixed point $p\in\mcP$) the $L$-th level iterate towards the IFS fractal $F_\mcT$.
\end{dfn}

We end this section by showing a lemma which will prove unexpectedly useful in Section \ref{s0305}.

\begin{lem} \label{s02040204} \textup{(Slope Lemma)}
The complex slope of the map $T_a,\ a\in\mcAf$ is the constant $\cf_a\in\C$, meaning
\[ \frac{T_a(z_1)-T_a(z_2)}{z_1-z_2} = \cf_a\ \ for\ any\ distinct\ \ z_{1,2}\in\C. \]
\end{lem}\begin{proof}
We show the property by induction with respect to $|a|$. For any $|a|=1$ address, i.e. $k\in\mcN$
\[ T_k(z_1)-T_k(z_2) = p_k + \cf_k(z_1-p_k) - p_k - \cf_k(z_2-p_k) = \cf_k(z_1-z_2). \]
Now let us suppose the property holds for $|a|\leq L$ and we show it for length $L+1$. Taking any $k\in\mcN$ we need the property for $(k,a)$.
\[ T_{(k,a)}(z_1)-T_{(k,a)}(z_2) = T_k(T_a(z_1))-T_k(T_a(z_2)) = \cf_k (T_a(z_1)-T_a(z_2)) = \]
\[ = \cf_k\cf_a(z_1-z_2) = \cf_{(k,a)}(z_1-z_2). \]
\end{proof}

\subsection{Optimal Transport over Measures} \label{s0205}

Transportation Theory, also referred to as the theory of Optimal Transport, is the study of the optimal transportation and allocation of resources. Given two finite sets of producers and consumers equal in number, with each producer supplying a specific consumer according to a bivariate cost function for transportation, the goal is to find the minimal arrangement between producers and consumers, called the optimal transport plan or map.

The discrete problem can be generalized to probability measures - the Monge-Kantorovich formulation, named after its two most influential creators. Like many optimization problems, this minimization problem has a dual maximization equivalent. When the cost function over a compact metric space $(X, d_X)$ is the metric itself, then the dual becomes the first Wasserstein distance (also called the Hutchinson metric) of two elements of the set of Borel probability measures $\nu_{1,2}\in\MX$ over $X$
\[ d(\nu_1, \nu_2) = \sup \left\{\int_X f\ \md(\nu_1-\nu_2): f\in\Lip \right\} \]
where $\Lip$ denotes the Lipschitz-1 functions over $X$. That this is indeed the dual to the minimization problem, was shown by Kantorovich and Rubinstein \cite{ba00041}. Furthermore $(\MX, d)$ is a complete metric space \cite{bb00007}.

Generally the transport plan is a pushforward map (transfer operator) of the form $T^*\nu = \nu\co\T^{-1}\ (\nu\in\MX)$ where $T:X\rightarrow X$. If $T^*$ is contractive in $d$, then by the Banach Fixed Point Theorem it has a unique invariant measure $\nub = T^*\nub$. Taking a finite weighted average of such pushforward operators, the resulting transfer operator
\[ \T = w_1 T_1^* +\ldots +w_n T_n^*\ \ \ \mathrm{where}\ \ \ w_k\in (0,1),\ w_1+\ldots+w_n =1 \]
will also be contractive in $d$ over $\MX$, so it possesses a fixed point too. Thus the following theorems hold for an IFS $\mcT=\{T_1,\ldots,\T_n\}$ in the form $T_k(z)=p_k+\vl_k R_k(z-p_k),\ z\in\R^2$ supposing that $\langle\mcT\rangle\subset X\subset \R^2$.

\begin{thm} \label{s02050001} \textup{(Hutchinson \cite{ba00007})}
For any IFS $\mcT=\{T_1,\ldots,T_n\}$ and weights $w_1,\ldots,w_n\in (0,1)$, the transfer operator $\T = w_1 T_1^* +\ldots +w_n T_n^*$ is contractive on $(\MX,d)$ and attains a unique invariant probability measure $\nub = \T\nub$ over the compact set $X\subset\R^2$, with support $\langle T_1,\ldots,T_n\rangle$. We call this the invariant measure with respect to the IFS $\mcT$. For any $\nu_0\in\MX$ the recursion $\nu_{L+1}=w_1 T_1^*\nu_L +\ldots +w_n T_n^*\nu_L\in\MX,\ L\in\N$ converges to the invariant measure.
\end{thm}

Note that with the primary fixed point $p$ and the notations of Section \ref{s020402}, the measures
\[ \nu_0 = \delta_p,\ \ \nu_L = \sum_{|a|=L} w_a T_a^*\delta_p = \sum_{|a|=L} w_a \delta_{T_a(p)}\in\MX,\ L\in\N \]
satisfy the above recursion, and so $\nub(S)=\lim (\nu_L)$ \cite{ba00099}. Also observe that the support of $\nu_L$ is concentrated on the $L$-th iterate $F_L:= H^L(\{p\})=\{T_a(p): a\in\mcA, |a|=L\}$ approaching the compact IFS fractal $F=\langle T_1,\ldots,T_n\rangle$.

Defining the following sets for a compact set $X\subset\R^2$
\[ \DX:= \left\{\vr\in C_{ae}(\R^2,\R): \vr\geq 0,\ \msupp\ \vr\subset X,\ \int_X\vr =1\right\} \]
\[ \PX:= \left\{\nu\in\MX:\ \exists\vr\in\DX\ \forall S\subset X\ \mathrm{Borel\ set}: \nu(S)=\int_S\vr\ \right\} \]
and keeping the above metric $d$, then $\Conv\{\delta_p: p\in\R^2\}\subset\PX\subset\MX$ and $\nub\in\PX$, so it remains true that there exists a unique invariant measure in $(\PX, d)$ with respect to $\mcT$, and it is the same as the earlier one. Furthermore $\T:\PX\frar\PX$ since $\msupp(\T\nu)=\mH(\msupp\ \nu)\subset X$ if $\nu\in\PX$ and $\langle\mcT\rangle\subset X$.

The study of transfer operators - also called Ruelle(-Perron-Frobenius-Markov) operators - is a rich field, and their largest eigenvalue is typically one, while their eigenfunctions are usually fractalline or self-similar in some sense. This has profound implications for classical mechanics, such as the increase of entropy or the irreversibility of time. An eigenvalue of one corresponds to a state of equilibrium.

For an introduction to Optimal Transport, see Villani \cite{bb00007} and McCann \cite{bc00009}, and for an overview of IFS, see \cite{ba00066, ba00007, ba00017, bb00009}. The current standard model for examining Poisson's Equation over IFS fractals is Kigami's formulation \cite{ba00049, ba00050, bb00010}, reviewed and further developed by Strichartz \cite{ba00051} and other researchers. For a numerical model of fluid dynamical advection via transfer operators, see the work of Froyland et al. \cite{ba00098}.

\subsection{Remarks} \label{s0206}

Some initial remarks are hereby made, which are to shape our approach to be presented in Section \ref{s03}. First of all, we reason that modeling the phenomenon of FDT via the Navier-Stokes Equations may well be a futile attempt. The discussed energy bursts fuelling the erratic characteristics of intermittency, are clear signs of the non-smooth nature of the evolution towards FDT. The Navier-Stokes Equations however, require temporal differentiability of the velocity field, which does not hold by the observed phenomenon of intermittent evolution, summarized in Section \ref{s020101}. Therefore one has no choice mathematically, but to abandon the Navier-Stokes Equations, in hope of a more fitting model for the evolution, in line with the experimental picture of the phenomenon. Leray \cite{ba00089} has argued the breakdown of the validity of the Navier-Stokes Equations in turbulence \cite{ba00091}, while Frisch et al. \cite{ba00077} have expressed the need for a genuine dynamical theory, as the obvious symmetries and conservation laws of the Navier-Stokes Equations seem insufficient \cite{ba00081}. On-off intermittency in turbulence discovered experimentally by Batchelor and Townsend \cite{ba00092} was placed in the context of failing classical determinism by Sommerer \cite{ba00040}. 

Secondly, it seems reasonable to raise the doubt whether Fully Developed Turbulence in full three dimensions even exists. Certainly, flows can approach above $10,000$ Reynolds numbers, but it is doubtable whether the other defining characteristics of FDT, including the equilibrium state of a stationary velocity field at $Re = \infty$ that is independent of viscosity, can evolve in a fully three-dimensional flow field. It seems that three-dimensional space has one too many degrees of freedom, while a limiting two-dimensional free surface may force the fluid to self-organize via self-similarity. Thus we restrict our attention to eddy interaction on the free surface in our effort to model FDT, inspired by the principles of Chaotic Advection.

Lastly, note that the standard mathematics generally employed in the theory of Chaotic Advection have not been detailed here. This is partly because the author finds the existing methods to be suboptimal for actualizing the potential in its concepts, and partly because the theory in its current form is often intuitive and in need of a rigorous mathematical language. The author feels that it needs a reformulation unifying the Aref \cite{ba00034, ba00035} and Ott \cite{ba00038} directions, in the spirit of Wiggins \cite{ba00043}.
\section{Fractal Potential Flows} \label{s03}

The main theoretical elements have been introduced which are to be woven together in this section and the next, with the intention to create an idealized model for Fully Developed Turbulence and the evolution towards it. Fractal Potential Flows are hereby introduced as a model, which arise due to the intermittent interaction of eddies. A weighted transfer operator is defined over a certain class of stream functions, and the existence of its unique fixed point - the invariant flow - is shown, characterized by a stream function whose rotated gradient field will serve as our model. The set of sink singularities of this flow field is shown to be an IFS fractal, explicitly resolving Mandelbrot's Conjecture \ref{s02010301}.

\subsection{Eddy Invariance} \label{s0301}

In Section \ref{s0202} we defined the stream function of an eddy with sink strength $q = C\ln\vl$ and vortex strength or circulation $\Gamma=C\vt$ as
\[ \psi(z) = -\frac{C\ln\vl}{2\pi}\Arg(z-p) + \frac{C\vt}{2\pi}\ln |z-p|\ \ \mmod\ -C\ln\vl \]
where $\cf\in\C,\ \vl=|\cf|<1,\ \vt=\Arg\ \cf$ and $C>0$ is an arbitrary parameter. The associated velocity field (continuously extended to the entire plane) becomes
\[ v[\psi](z):=\frac{C\Log\ \cf}{2\pi}\frac{1}{|z-p|}\frac{z-p}{|z-p|}. \]
Notice that the eddy stream function is invariant under the contraction $T(z)=p+\cf(z-p)$, meaning $T^*\psi =\psi$. This is expected geometrically, since the orbit $T^t(z_0)=p+\cf^t(z_0-p),\ t\in\R$ traces out a logarithmic spiral centered at $p\in\C$, and the streamlines of constant $\psi$ are spirals of the same pitch. So in other words, $\psi$ is the fixed point of the pushforward transfer operator $T^*$. This raises the question whether $\psi$ is unique, and if not, then what space would guarantee its uniqueness. Defining $\psi_0(z):=\Arg(z-p)\ \mmod\ \vt$ we see that $T^*\psi_0 = \psi_0$ also holds, therefore $\psi$ cannot be unique in such a general setting. So our ultimate goal becomes finding the proper function space where uniqueness can be guaranteed, with a reasonable physical interpretation. Let us begin our search by deriving some fundamental properties of transfer, which are to have profound implications.

\begin{thm} \label{s03010001}
For a similarity contraction of the form $T(z)=p+\cf(z-p),\ |\cf|<1$ the following properties hold (with differentiation in the $\R^2$ sense, and $v[\psi]$ being the continuous extension of $\vi\grd\psi$ to the plane).
\[ \lap T^* = \frac{1}{\vl^2}\ T^*\!\lap\ \ \ and\ \ \ \widetilde{T^*\psi}\equiv T^*\tilde{\psi}\ \ \ and\ \ \ v[T^*\psi] = \frac{\cf}{|\cf|^2}\ T^{*}v[\psi]. \]
\end{thm}\begin{proof}
We show the above in the $\R^2$ sense first, which translates to the complex sense, using $T(z)=p+\vl R(z-p),\ p,z\in\R^2$ where $R\in\R^{2\times 2}$ is the rotation matrix corresponding to $\me^{\vt\vi}$. We first show the third property, keeping in mind that $v[\psi](z)=i\grd\psi(z)\ \mae$ Gradient is a column vector in $\R^2$, which corresponds to the complex vector $\partial_1+\vi\partial_2$. Differentiation $\D$ results in a row vector however, so we must take a transpose, meaning $\grd=\D^T$. Applying the generalized Chain Rule \cite{bb00019}, we get
\[ \grd(\psi\co T^{-1}) = \D(\psi\co T^{-1})^T = \left((\D\psi\co T^{-1})\cdot\frac{1}{\vl}R^{-1}\right)^T = \frac{1}{\vl}R\cdot\grd\psi\co T^{-1}. \]
Multiplying by $\vi$ ($90^\circ$ rotation), we get the third property. To get the first property, we observe that $\lap = \tr\ \D\grd$, so by the Chain Rule, the properties of trace, and the above equation, we have
\[ \lap T^*\psi = \tr\ \D\grd T^*\psi = \tr\ \D\left(\frac{1}{\vl}R\cdot T^*\grd\psi\right) = \tr\left(\frac{1}{\vl}R\cdot T^*\D\grd\psi\cdot\frac{1}{\vl}R^{-1}\right) = \frac{1}{\vl^2}T^*\lap\psi. \]
The second property also follows from the third, observing that in the above complex sense $\psi$ and $\tilde{\psi}$ are conjugates iff $\grd\psi=\vi\grd\tilde{\psi}$ by Section \ref{s0202}. So taking $T^*$ of both sides, multiplying by $\vi$ and dividing by $\vl$, we get that $\grd(T^*\psi) = \vi\grd(T^*\tilde{\psi})$ implying that $\grd(T^*\tilde{\psi})=\grd(\widetilde{T^*\psi})$ and thus $\widetilde{T^*\psi}\equiv T^*\tilde{\psi}$.
\end{proof}

Among many things, the first property also implies that if $\psi$ is $\mae$ harmonic then so is $T^*\psi$. Together with the second property, this implies that the space of potential flows is closed under pushforward transfer by a similarity contraction.

The underlying idea above is that one application of $T^*$ to a flow field represented by the stream function, corresponds to one intermittent energy burst fuelling an eddy to transform the flow field along the logarithmic spiral orbits of $T$. If $T^*$ can be shown to be contractive over stream functions in some complete space, then the final equilibrium fixed point stream can also be shown.

\subsection{The Transfer Operator} \label{s0302}

Considering that potential flows form a vector space as discussed in Section \ref{s0202} and by Theorem \ref{s03010001}, the weighted transfer operator
\[ \T\psi = w_1 T_1^*\psi +\ldots + w_n T_n^*\psi \]
for any IFS $\{T_1,\ldots,\T_n\},\ T_k(z)=p_k+\cf_k(z-p_k)$ and weights $w_k\in(0,1),\ \sum_k w_k=1$, maps the stream function of a potential flow to another such flow. One application of $\T$ to a flow field corresponds to the weighted application of $T_k^*\in\{T_1^*,\ldots, T_n^*\}$ with probability $\mP(\T\psi(z_0) = T_k^*\psi(z_0))=w_k$ at some point $z_0$ in the plane. Indeed, the iteration of the transfer operator $\T$ corresponds to the intermittent randomly alternating contractive action of a system of $n$ eddies, acting along the orbits of $T_k$. Here the flow field is piecewise steady in between the energy bursts fuelling a particular eddy at a certain intermittent moment. During the steady periods, the flow field is of the form $\T^L\psi_0,\ L\in\N$ for some initial flow $\psi_0$. The partitioning intermittent moments in time $t_L$ can be spaced in any way over the timeline. This constitutes our mathematical reformulation of the Blinking Vortex-Sink System of Chaotic Advection in Section \ref{s020301}. (See Section \ref{s0404} for a more rigorous mathematical formulation of the above via conditional probabilities, as well as details on experimental verification.)

Observe that after the $L$-th iteration we have
\[ \psi_L = \T^L\psi_0 = \sum_{|a|=L} w_a\ T_a^*\psi_0 \]
where $T_a$ and $w_a$ were defined in Definition \ref{s02040201}. Probabilistically speaking, this implies that $\psi_L$ is the stochastic superposition of $n^L$ eddies with transfers $T_a^*$, transforming the flow field at the $L$-th intermittent moment $t_L$ with probabilities $\mP(\psi_L(z_0) = T_a^*\psi_0(z_0))=w_a$. Also note that by Theorem \ref{s03010001} the corresponding velocity transfer operator is
\[ v[\T\psi]= w_1\frac{\cf_1}{|\cf_1|^2}\ T_1^*v[\psi] +\ldots + w_n\frac{\cf_n}{|\cf_n|^2}\ T_n^*v[\psi]. \]
There are two possible interpretations of the weighted transfer iteration - probabilistic and physical. Probabilistically it means that considering the next iterate $\psi_{L+1}(z_0)$ to be a random variable located at each $z_0\in\C$, its expected value is $\T\psi_L(z_0)$. Physically, weighted transfer can be interpreted as weighting the effect each eddy - represented by $T_k^*$ - has on the fluid at an intermittent moment at a particular point in space. This is essentially a chaos game.

Succinctly, we claim that iteration corresponds to intermittency, and hereby theorize that the intermittent evolution of turbulence towards the equilibrium state of Fully Developed Turbulence, corresponds to the iteration of $\T$ ad infinitum to an invariant flow $\T\psib = \psib$. Therefore $\psib$ will represent our model for the fully developed turbulent flow field, and showing its unique existence now becomes our primary goal, along with finding an appropriate function space where this is possible and physically most reasonable. Progressing towards this goal, we state a weighted generalization of Theorem \ref{s03010001}.

\begin{thm} \label{s03020001}
For the weighted transfer operator\ $\T$\ the following properties hold.
\[ \lap\T = \sum_{k=1}^n \frac{w_k}{\vl_k^2}\ T_k^*\lap\ \ \ and\ \ \ \widetilde{\T\psi}\equiv \T\tilde{\psi}\ \ \ and\ \ \ v[\T\psi] = \sum_{k=1}^n w_k \frac{\cf_k}{|\cf_k|^2}\ T_k^* v[\psi]. \]
\end{thm}\begin{proof}
The proof follows from Theorem \ref{s03010001}. Note that the first property implies the preservation of harmonicity $\mae$ or physicality upon transfer, as discussed in Section \ref{s0202}.
\end{proof}

\subsection{Flow Character} \label{s0303}

In constructing the proper flow space, our inspiring objective is to ensure the unique existence of an eddy as the attractor of its generating pushforward transfer map. In resolving this question, we show a correspondence between eddies and the Dirac delta function, which hints at a possible general correspondence between stream functions and the density functions of probability measures, via Poisson's Equation. Finding the ideal type of boundary condition will prove crucial to our quest, and meanwhile the proper space of flows shall gradually reveal itself.

Firstly notice that the velocity field of the eddy discussed in Section \ref{s0301} is characterized by the complex parameter $c=\frac{C}{2\pi}\Log\ \cf$ which can be extracted from the stream function $\psi$ by the operation
\[ v[\psi](z)\ \overline{z-p} = c\ \frac{z-p}{|z-p|^2}\ \overline{z-p} = c. \]
We will denote the stream function of an eddy with character $c\in\C,\ \mRe\ c<0$ and centered at $p\in\C$, with $\qc:=-2\pi\mRe\ c,\ \gc:=2\pi\mIm\ c$ as
\[ \psi_{c,p}(z) := \frac{\qc}{2\pi}\Arg (z-p) + \frac{\gc}{2\pi}\ln |z-p|\ \ \mmod\ \qc\ \ \ \mathrm{and}\ \ \ \psi_c:=\psi_{c,0}. \]
Certain flows may have a similar ``eddy character'' when their flow field is zoomed out, even if they exhibit varying streamline behaviour locally around the origin, as defined below.

\begin{dfn} \label{s03030001}
We say that a stream function $\psi:\C\frar\R$ satisfies the boundary condition at infinity with character $c\in\C$, if for any $p\in\C$ and any sequence $(z_j)\subset\C, |z_j-p|\frar\infty$ we have $\exists\lim_{j\frar\infty} v[\psi](z_j)\ \overline{z_j-p} = c$ where the limit is taken in the complex sense. Let this property be denoted as\ $\psi\in\BCci$\ or\ $\chr(\psi)=c$.
\end{dfn}

Remark that character is independent of the choice of $p\in\C$. To see this, take
\[ \lim_{j\frar\infty} |v[\psi](z_j)| = \lim_{j\frar\infty} \left|\frac{v[\psi](z_j)\ \overline{z_j-p}}{\overline{z_j-p}}\right| = \lim_{j\frar\infty} \frac{|c|}{|z_j-p|} = 0 \]
so $\lim (v[\psi](z_j))=0$. Taking some other $p'\in\C$ we have for the above $(z_j)$ that $|z_j-p'|\frar\infty$ since $|z_j-p|\leq |z_j-p'|+|p'-p|$ and that
\[ v[\psi](z_j)\ \overline{z_j-p'} = v[\psi](z_j)\ \overline{z_j-p} + v[\psi](z_j)\ \overline{p-p'}\ \frar\ c+0= c\ \ \mathrm{as}\ \ j\frar\infty. \]
In what follows, let us denote with $C_{ae}^2(\R^2)$ the $\mae$ defined and twice $\mae$ continuously differentiable functions $\psi$, for which $\vi\grd\psi$ can be continuously extended to the entire plane, denoted as $v[\psi]$, and furthermore the set of singularities $S_{\psi}$ of $v[\psi]$ is compact and nonempty. Let $L^\infty(\R^2)$ denote the set of bounded functions $\psi$ over the plane, for which the following also holds
\[ \exists R,C>0\ \forall z\in\C\ (|z|\geq R):\ |v[\psi](z)|\ |z|\leq C. \]
Note that for such $z$ we also have $|v[\psi](z)|\leq C/R$. Clearly $C_{ae}^2(\R^2)\cap L^\infty(\R^2)$ is a vector space.

\begin{thm} \label{s03030002}
Over $C_{ae}^2(\R^2)\cap L^\infty(\R^2)$, character is a linear map that is invariant under weighted transfer, or convolution with a density function of bounded support over the plane.
\end{thm}\begin{proof}
The linearity of $\chr$ is trivial by definition. Let us suppose that $\chr(\psi)=c$. Since for some $p\in\C$ and $(z_j)\subset\C,\ |z_j-p|\frar\infty$ sequence $z_j-p = \cf_k(T_k^{-1}(z_j)-p_k)+(p_k-p)$ and applying Theorem \ref{s03020001} we have that
\[ v[\T\psi](z_j)\ \overline{z_j-p} = \sum_{k=1}^n w_k\frac{\cf_k}{|\cf_k|^2}\ v[\psi](T_k^{-1}(z_j))\ \overline{\cf_k(T_k^{-1}(z_j)-p_k)+(p_k-p)} = \]
\[ = \sum_{k=1}^n w_k\frac{\cf_k\bar{\cf}_k}{|\cf_k|^2}\ v[\psi](T_k^{-1}(z_j))\ \overline{T_k^{-1}(z_j)-p_k} + \sum_{k=1}^n w_k\frac{\cf_k}{|\cf_k|^2}\ v[\psi](T_k^{-1}(z_j))\ \overline{p_k-p} \]
which approaches $\sum_k w_k c + \sum_k 0 = c$ as $j\frar\infty$ implying that $\chr(\T\psi)=c$.\\
Now let $\psi\in C_{ae}^2(\R^2)\cap L^\infty(\R^2)$, let $\vr$ be a density function with bounded support $\msupp\ \vr$ over $\R^2$, let $p\in\C$ be fixed and arbitrary, and let $(z_j)\subset\C$ be again any sequence for which $|z_j-p|\frar\infty$. We employ scalar and complex vector Riemann integrals as follows.
\[ \grd(\psi\ast\vr)(z) = \int_{\R^2} \grd\psi(z-w)\vr(w)\ \md w \]
\[ v[\psi\ast\vr](z_j)\ \overline{z_j-p} = \int_{\R^2} v[\psi](z_j-w)\ \overline{(z_j-w)-(p-w)}\ \vr(w)\ \md w. \]
We show that the limit as $j\frar\infty$ can be taken inside the integral, using Lebesgue's Dominated Convergence Theorem. Since $\vr$ has bounded support, we choose $R'>0$ to be the radius of some closed ball $S=B(p,R'):=\{z\in\C:|z-p|\leq R'\}$ which contains $\msupp\ \vr$. With the appropriate $R,C>0$ for $\psi\in L^\infty(\R^2)$, we have for all $w\in S$ and $j\in\N\ (d(z_j,S)\geq R)$ that $|z_j-w|\geq R$ and
\[ |v[\psi](z_j-w)\ \overline{z_j-p}|\leq C \frac{|z_j-p|}{|z_j-w|} \leq C\left(1+\frac{|w-p|}{R}\right)\leq C\left(1+\frac{R'}{R}\right) \]
meaning the elements of the function sequence $w\mapsto v[\psi](z_j-w)\ \overline{z_j-p}\ (j\in\N)$ are uniformly bounded over $S$ (for large enough indices $j\in\N$ such that $d(z_j,S)\geq R$ which are guaranteed by $|z_j-p|\tti$). So using the aforementioned theorem and the above properties
\[ \lim_{j\tti} v[\psi\ast\vr](z_j)\ \overline{z_j-p} = \lim_{j\tti} \int_S v[\psi](z_j-w)\ \overline{(z_j-w)-(p-w)}\ \vr(w)\ \md w = \]
\[ = \int_S\ \lim_{j\tti} v[\psi](z_j-w)\ \overline{(z_j-w)-(p-w)}\ \vr(w)\ \md w = \int_S c \vr(w)\ \md w = c \]
where the latter limit is taken pointwise. This implies that $\psi\ast\vr$ preserves the character $c$.
\end{proof}

\begin{thm} \label{s03030003}
For any density function $\vr$ of bounded support over the plane, and any $c\in\C,\ \mIm\ c\neq 0$, there exists a $\psi\in C_{ae}^2(\R^2)\cap L^\infty(\R^2)\cap BC_c^\infty$ for which $\lap\psi = \gc\vr\ \mae$ Furthermore, this $\psi$ is unique up to a gradient $\mae$ and $\psi\equiv\psi_c\ast\vr$.
\end{thm}\begin{proof}
The existence of such a $\psi$ is guaranteed using the Green function method of convolving the fundamental solution $\psi_c$ ($\lap\psi_c = \gc\delta_0\ \mae$) with $\vr$ which will also have a character of $c$ by the previous theorem, and will also be bounded by trivial calculation. Note that convolution is known to preserve differentiability as many times as its terms are differentiable in total.\\
Let us proceed to the question of uniqueness, and suppose indirectly that $\exists\psi_{1,2}\in\BCci,\ \psi_1\not\equiv\psi_2$ such that $\lap\psi_1 = \gc\vr = \lap\psi_2\ \mae$ Then $\lap\psi = 0\ \mae$ with $\psi:=\psi_1-\psi_2$. Applying the Divergence Theorem for any compact set $S\subset\R^2$ with piecewise smooth boundary
\[ \int_S |\grd\psi|^2 = \int_S |\grd\psi|^2+\psi\lap\psi = \int_S \mdiv(\psi\grd\psi) = \int_{\partial S} \psi\langle\grd\psi, u\rangle. \]
Since $\psi_{1,2}$ are bounded, then so is $\psi$ with say $|\psi|\leq M$ over the plane.\\
So for any sequence of disks $S_j:=B(p,r_j),\ 0<r_j<r_{j+1},\ j\in\N,\ \lim (r_j)=\infty$ for which the set of singularities $S_\psi$ satisfies $S_{\psi}\subset S_1,\ \partial S_{\psi}\cap\partial S_1=\mes$, we have
\[ \int_{S_j} |\grd\psi|^2 = \left|\int_{\partial S_j} \psi \langle\grd\psi, u\rangle\right|\leq M\!\int_{\partial S_j} \left|\left\langle\grd\psi(z), \frac{z-p}{|z-p|}\right\rangle\right|\md z \leq M\!\int_{\partial S_j} |v[\psi](z)|\ \md z \]
where the integrals are taken in the $\R^2$ sense. If we can show that this last integral vanishes as $j\frar\infty$, then $\int_{\R^2} |\grd\psi|^2 =0$ and so $\grd\psi = 0\ \mae$ which would contradict that $\psi_1\not\equiv\psi_2$.\\
Take any $z_j\in\argmax_{z\in\partial S_j} |v[\psi](z)|,\ j\in\N$ which are nonempty sets by the Extreme Value Theorem, considering the compactness of $\partial S_j$ and the continuity of $|v[\psi]|$ on $\partial S_j\subset\R^2\setminus S_{\psi}$. Since $z_j\in\partial S_j,\ j\in\N$ we have $|z_j-p|=r_j\frar\infty$ as $j\frar\infty$ which by $\chr(\psi_{1,2})=c$ implies that $v[\psi](z_j)\ \overline{z_j-p}\frar 0$ so
\[ \int_{\partial S_j} |v[\psi](z)|\ \md z \leq \int_{\partial S_j} |v[\psi](z_j)|\ \md z = |v[\psi](z_j)\ \overline{z_j-p}|\ \frac{1}{|z_j-p|}\ 2\pi r_j\ \frar\ 0\ \ \mathrm{as}\ \ j\frar\infty. \]
\end{proof}

By the above results, we may conclude that the eddy $\psi_{c,p}$ uniquely corresponds via Poisson's Equation to the Dirac delta function $\delta_p$ among all $L^\infty(\R^2)$ functions of character $c$, where the uniqueness is guaranteed up to a gradient.

\subsection{The Flow Space} \label{s0304}

\begin{dfn} \label{s03040001}
Let $X\subset\R^2$ be a compact set that contains the IFS fractal $F=\langle T_1,\ldots,T_n\rangle$ with $\mleb(F)=0$. Let $c\in\C$ be some number for which $\mRe\ c <0,\ \mIm\ c\neq 0$. Then define for $\psi\in C_{ae}^2(\R^2)$
\[ \mu[\psi](S)=\mu_c[\psi](S):=\frac{1}{\gc}\int_S\lap\psi\ \ \ (S\subset\R^2). \]
Let the set of flows $\psi$ over $X$ with character $c$ be defined as
\[ \Psi_{X,c}:= \{\psi\in C_{ae}^2(\R^2)\cap L^\infty(\R^2)\cap BC_c^\infty:\ \Ran\ \psi\subset [0,\qc),\ \mu[\psi]\in\PX \}. \]
We refer to $\mu:\Psi_{X,c}\frar\PX$ as the measure map, and to $\mu[\psi]$ as the measure induced by $\psi$.\\
Recalling the metric $d$ from Section \ref{s0205}, for any $\psi_{1,2}\in\PsXc$ denote
\[ D(\psi_1,\psi_2)=D_{X,c}(\psi_1,\psi_2):=d(\mu[\psi_1],\mu[\psi_2])=\sup\left\{\frac{1}{\gc}\int_X f\ \lap(\psi_1-\psi_2): f\in\Lip\right\}. \]
As in Section \ref{s0202}, we consider two functions to be congruent $\psi_1\equiv\psi_2$ iff $\grd\psi_1=\grd\psi_2\ \mae$ over $X$. Let the equivalence class of $\psi$ be denoted as $[\psi]:=\{\psi_0\in\Psi_{X,c}: \psi\equiv\psi_0\}$ and let
\[ \left[\Psi_{X,c}\right]:=\{[\psi]: \psi\in\Psi_{X,c}\} = \Psi_{X,c}/\Ker(\mu_c). \]
Further denote $D([\psi_1], [\psi_2])=D_{X,c}([\psi_1], [\psi_2]):= D_{X,c}(\psi_1,\psi_2)$.\\
Then we call $([\Psi_{X,c}], D)$ the flow space over $X$ with character $c$.
\end{dfn}

\begin{thm} \label{s03040002}
$([\Psi_{X,c}], D)$ is a well-defined metric space. The measure map $\mu:[\Psi_{X,c}]\frar\PX$ is an isometric isomorphism, and the weighted transfer operator preserves flows $\T:\Psi_{X,c}\frar\Psi_{X,c}$. Furthermore, the two maps commute.
\end{thm}\begin{proof}
First of all, let us observe that $\mu:\bPsXc\frar\PX$ is clearly an isometry, since by definition $d(\mu[\psi_1],\mu[\psi_2])=D([\psi_1],[\psi_2])$. Also $\mu$ is bijective iff $\forall\nu\in\PX\ \exists! [\psi]\in\bPsXc:\ \mu[\psi]=\nu$ and the equality $\mu[\psi]=\nu$ here means that with the density function $\vr$ belonging to $\nu\in\PX$ we need
\[ \mu[\psi](S)=\frac{1}{\gc}\int_S \lap\psi = \nu(S) = \int_S\vr\ \ \ \forall S\subset X\ \mathrm{Borel\ sets} \]
which is equivalent to $\lap\psi = \gc\vr\ \mae$ on $X$ by the du Bois-Reymond Lemma. The unique existence of such a $\psi\in\PsXc$ for any $\vr$ was shown in Theorem \ref{s03030003} up to a gradient $\mae$, meaning up to a congruence or equivalence class. So we have that $\mu$ is a bijective isometry, or an isometric isomorphism.\\
For the well-definition of the space, we must show that $D$ is a metric over the set \bPsXc. $D$ is positive, symmetric, and inherits the triangle inequality from $d$. The only question that remains is whether $D([\psi_1],[\psi_2])=0$ implies $[\psi_1]=[\psi_2]$. Since $0=D([\psi_1],[\psi_2])=d(\mu[\psi_1],\mu[\psi_2])$ we have that $\mu[\psi_1]=\mu[\psi_2]$, and since $\mu$ has been shown to be bijective, this implies that $[\psi_1]=[\psi_2]$ or $\psi_1\equiv\psi_2$.\\
We proceed to proving that $\T$ and $\mu$ commute, meaning $\T\mu = \mu\T$. Recall from Theorem \ref{s03020001} that
\[ \lap(\T\psi)(z) = \sum_{k=1}^n \frac{w_k}{\vl_k^2}\ (\lap\psi)\co T_k^{-1}(z) \]
so for any $S\subset X\subset \R^2$ and $k\in\{1,\ldots,n\}$ taking Riemann integrals over $\R^2$ we have
\[ \mu[T_k^*\psi](S) = \frac{1}{\gc}\int_S\lap(\psi\co T_k^{-1}) = \frac{1}{\gc}\int_S \frac{1}{\vl_k^2}(\lap\psi)(T_k^{-1}(z))\ \md z = \]
\[ = \frac{1}{\gc}\int_{T_k^{-1}(S)}\lap\psi(w)\ \md w = \mu[\psi](T_k^{-1}(S)) = (T_k^*\mu[\psi])(S) \]
so by the linearity of $\mu$ we have that $\mu[\T\psi]=\T\mu[\psi],\ \psi\in\PsXc$ or succinctly $\mu\T = \T\mu$.\\
Lastly, we show that $\T$ maps from $\PsXc$ to itself. $\T$ preserves $C_{ae}^2(\R^2)$ since $S_{\T\psi}=\mH(S_{\psi})$, and it also preserves character according to Theorem \ref{s03030002}. Since $\T$ is a convex combination, it preserves that $\Ran(\T\psi)\subset[0,\qc)$ and $\T\psi\in L^\infty(\R^2)$ since the latter is a vector space. Furthermore, we have that $\mu[\T\psi]=\T\mu[\psi]\in\PX$ due to $\mu[\psi]\in\PX$ and $\T:\PX\frar\PX$ by Section \ref{s0205}.
\end{proof}

Physically speaking, the above space $([\Psi_{X,c}], D)$ represents the fact that the intermittent interaction of a system of $n$ eddies, can be considered in a locally isolated manner over a compact $X\subset\R^2$ of reasonable scale. $\bPsXc$ versus $\PsXc$ signifies that it is only the velocity field induced by a stream function that matters. Prior to the start of their interaction, the eddies $\psi_{c_k,p_k}$ (notation of Section \ref{s0303}) superpose initially in a flow with a certain character $c$ as follows. Letting $\psi_0:=\sum_k w_k T_k^*\psi_{c_k,p_k}$ then by Theorem \ref{s03010001}
\[ v[\psi_0]= \sum_{k=1}^n w_k\frac{\cf_k}{|\cf_k|^2}\ T_k^*v[\psi_{c_k,p_k}] = \sum_{k=1}^n w_k\ v[\psi_{c_k,p_k}] \]
so clearly $c=\chr(\psi_0)=\sum_k w_k c_k$. As the intermittent interaction of the eddies progresses according to the transfer operator $\T$, their overall character will be preserved to be this $c$ by Theorem \ref{s03030002}.

\subsection{The Invariant Flow} \label{s0305}

\begin{thm} \label{s03050001}
The transfer operator $\T$ is contractive over $(\bPsXc, D)$, so any $(\psi_L)\subset\PsXc$ sequence satisfying $\psi_{L+1}\equiv\T\psi_L\in\PsXc$ converges in the metric $D$ to a $\psib\in\PsXc$ unique up to a gradient $\mae$ Furthermore
\[ \psib\equiv\psi_c\ast\nub \equiv \lim_{L\frar\infty} \sum_{|a|=L} w_a\ \psi_{c, T_a(p)}\in\PsXc \]
and $\T\psib\equiv\psib$ where $\nub$ is the invariant measure of Theorem \ref{s02050001} and $p$ is the primary fixed point. We call this $\psib$ the invariant flow (or fractal potential flow) with respect to the IFS $\mcT=\{T_1,\ldots,T_n\}$, the corresponding weights $w_1,\ldots,w_n\in (0,1)$, the compact set $X$, and the character $c\in\C$.
\end{thm}\begin{proof}
The contractivity of transfer over stream functions follows directly from that of the measure transfer in Theorem \ref{s02050001}, considering that the measure map is isometric by Definition \ref{s03040001}.\\
At this point one would expect to show the completeness of $(\bPsXc,D)$ in order to conclude the unique existence of a fixed flow of $\T$ by the Banach Fixed Point Theorem. This however becomes unnecessary, considering that we have a powerful isometric isomorphism $\mu:[\Psi_{X,c}]\frar\PX$ between flows and probability measures, which commutes with $\T$ by Theorem \ref{s03040002}.\\
By Theorem \ref{s02050001} and the remarks made on it, there exists a unique $\nub\in\PX$ such that $\T\nub=\nub$. Furthermore, by Theorem \ref{s03040002} there exists a unique $[\psib]\in\bPsXc$ such that $\mu[\psib]=\nub$ and
\[ \mu[\T\psib] = \T\mu[\psib] = \T\nub = \nub = \mu[\psib]\ \rar\ \T\psib\equiv\psib \]
since $\mu$ is bijective. Now let us suppose indirectly that $[\psib]$ is not a unique fixed point, meaning there is another $\psi\not\equiv\psib$ for which $\T\psi\equiv\psi$. Then
\[ \mu[\psi] = \mu[\T\psi] = \T\mu[\psi]\ \rar\ \mu[\psi]=\nub=\mu[\psib]\ \rar\ \psi\equiv\psib \]
which is a contradiction.\\
Now consider any sequence $(\psi_L)\subset\PsXc$ satisfying $\psi_{L+1}\equiv\T\psi_L$ and define $\nu_L:=\mu[\psi_L]$. Then $\nu_{L+1}=\mu[\T\psi_L]=\T\nu_L$ so by Theorem \ref{s02050001} we have that $\exists\lim (\nu_L)=\nub$ in $(\PX,d)$. Since $D([\ \cdot\ ],[\ \cdot\ ])=d(\mu[\ \cdot\ ],\mu[\ \cdot\ ])$ and $\mu$ is injective, we have that $\exists\lim ([\psi_L])=[\psib]$ or $\exists\lim (\psi_L)\equiv\psib$ where the limit is taken in the metric $D$.\\
Clearly $\psi_c\in\PsXc$ and it is a fundamental solution $\lap\psi_c= \gc\delta_0$ so by $\mu[\psib]=\nub$ from the proof of Theorems \ref{s03030003} and \ref{s03040002} we see that $\psib\equiv\psi_c\ast\nub$.\\
Define the sequence $\psi_0:=\psi_{c,p},\ \psi_L:=\sum_{|a|=L} w_a \psi_{c, T_a(p)}\in\PsXc,\ L\in\N$. The Slope Lemma \ref{s02040204} implies that $T_k^{-1}(z)-T_a(p)=\cf_k^{-1}(z-T_kT_a(p)),\ a\in\mcA$ and $\psi_c(\cf_k^{-1}w)\equiv\psi_c(w)-\psi_c(\cf_k)\ (\mmod\ \qc)$ so
\[ \T\psi_L(z)=\sum_{k=1}^n w_k \sum_{|a|=L} w_a \psi_c(T_k^{-1}(z)-T_a(p))\equiv \sum_{k=1}^n w_k \sum_{|a|=L} w_a \psi_c(z-T_kT_a(p)) = \psi_{L+1}(z). \]
Thus $(\psi_L)\subset\PsXc$ satisfies $\T\psi_L\equiv\psi_{L+1}$ so by the above $\exists\lim (\psi_L)\equiv\psib$.
\end{proof}

Examining the intriguing quantities $\psi_c(\cf_a)$ further by which $\psi_{c, T_a(p)}$ and $T_a^*\psi_{c,p}$ differ
\[ \psi_L(z)-\T^L\psi_0(z) \equiv \sum_{|a|=L} w_a\psi_c(\cf_a) \equiv \frac{\qc}{2\pi}\sum_{|a|=L} w_a\vt_a + \frac{\gc}{2\pi}\sum_{|a|=L} w_a\ln\vl_a\ \ (\mmod\ \qc) \]
and choosing the weights $w_k=\vl_k^s$ (where $s>0,\ \sum_k\vl_k^s=1$ is the so-called pseudo- or similarity dimension of the IFS fractal $\langle T_1,\ldots,T_n\rangle$ which corresponds to its Hausdorff dimension if the Open Set Condition holds \cite{ba00007}) we have as the factor of $\frac{\gc}{2\pi}$ above the following quantity
\[ -\frac{1}{s}\sum_{|a|=L} \vl_a^s\ln\vl_a^s =  -\frac{1}{s}\sum_{|a|=L} w_a\ln w_a. \]
Considering the weights $w_k=\vl_k^s$ to be probabilities, this is the Gibbs Entropy Formula for a collection of classical particles. It can be interpreted as the entropy of the set of points $F_L=\{T_a(p): a\in\mcA, |a|=L\}$ at each level $L\in\N$ in the Chaos Game evolution towards the IFS fractal attractor $F=\langle T_1,\ldots,T_n\rangle = \Cl\{T_a(p):a\in\mcA\}$. Remarkably $\frac{1}{s}$ plays the role of Boltzmann's constant, hinting at some potentially deeper interpretations.

\subsection{Equilibrium Points of the Invariant Flow} \label{s0306}

\subsubsection{Sink Singularities} \label{s030601}

From the representation shown for the invariant flow
\[ \psib(z) \equiv \lim_{L\frar\infty} \sum_{|a|=L} w_a\ \psi_{c, T_a(p)}(z) \equiv \lim_{L\frar\infty} \sum_{|a|=L} w_a\ \psi_c(z-T_a(p)) \]
where the limit is taken in the sense of $(\bPsXc, D)$, we see that $\psib$ is an infinite shifted superposition of the eddy $\psi_c$ with character $c$, preserved by $\psib$. Since $\psi_c$ is centered at the origin, each weighted eddy $w_a\psi_c(z-T_a(p))$ is centered at $T_a(p)$.

Therefore the set of sink singularities of $\psib$ is the IFS fractal $F=\langle T_1,\ldots,T_n\rangle=\Cl\{T_a(p):a\in\mcA\}$ where $p$ is the primary fixed point. Since as hereby reasoned $\psib$ represents a fully developed turbulent flow field, this resolves Mandelbrot's Conjecture \ref{s02010301}. Note that it was assumed in Definition \ref{s03040001} that $\mleb(F)=0$ to preserve $\mae$ differentiability. IFS fractals mostly have a non-integer Hausdorff dimension and thus zero Lebesgue measure, but in certain special cases, such as Dragon Curves, the dimension can equal two (to be resolved in Section \ref{s0307}).

Since the invariant flow concentrates on an IFS fractal, we must focus our efforts on the study of such sets, in order to uncover the characteristics of fractal potential flows (see \cite{ph00004}). It must be emphasized that by the proofs of Theorems \ref{s03010001} and \ref{s03020001}, having similarity contractions in the IFS is necessary for preserving the physicality of flows under transfer.

\subsubsection{Saddle Points} \label{s030602}

When visualizing $\T^L\psi_{c,p}$ for large enough $L\in\N$, one observes a thinning of the basins of attraction of each eddy center $F_L = \{T_a(p): a\in\mcA, |a|=L\}$ which raises the question whether the thinning continues on to a width of zero. The basins are partitioned by directrices (manifolds) of the saddle points (hyperbolic equilibria) of the flow field. In the neighbourhood of each partitioning infinite separatrix (stable manifold) belonging to these saddles (the unstable one ending in sinks), the flow behaviour becomes chaotic - meaning a tracer particle starting down the streamline near one side of the separatrix, may end up at a distant sink singularity, relative to if it had started near the other side (see Figs. \ref{s04010002} and \ref{s04010003}). When the flow field is considered as the phase portrait of a Hamiltonian system, then this signifies sensitivity to initial conditions near a particular separatrix. For the entire flow field to be considered chaotic, we must show that such partitioning separatrices ``cover'' the entire flow field, in some sense. We also intuit that this would be implied if the saddle points were shown to be dense in the IFS fractal $\langle T_1,\ldots,T_n\rangle$.

\begin{cnj} \label{s03060201}
The saddle points of $\psib$ are dense in $\langle T_1,\ldots,\T_n\rangle$, and the infinite separatrices of the saddle points are dense in the plane.
\end{cnj}

\subsection{Degenerate Cases} \label{s0307}

There are two possible ways the presented model can degenerate. The first is when $\gc=0$ in the definition of the measure map, seemingly preventing the rest of the discussion, from Definition \ref{s03040001} onwards. This case can be easily resolved however, and it is important since it models the planar electric field of an IFS fractal.

Let us split the initial stream function $\psi_0$ of character $c=\mRe\ c<0$ into the average of two functions $\psi_{1,2}$ with characters $c_{1,2}:=\chr(\psi_{1,2})=c\pm d\vi$ for some arbitrary $d\in\R\setminus\{0\}$. Note that such a representation of $\psi_0$ exists, since a $\psi_1$ of character $c+d\vi$ clearly exists, such as the eddy $\psi_{c+d\vi}$, and then $\psi_2:=2\psi_0-\psi_1$ will have a character of $c-d\vi$ and $\psi_0=\frac{\psi_1+\psi_2}{2}$.

Keeping the IFS, the weights, and $X$, by Theorem \ref{s03050001} the invariant flows $\psib_{1,2}\equiv\T\psib_{1,2}\in\Psi_{X,c_{1,2}}$ exist for both $c_{1,2}$ since $\mIm\ c_{1,2}=\pm d\neq 0$. Since $\equiv$ means equality in gradient $\mae$, we have for $\psib:=\frac{\psib_1+\psib_2}{2}$ that $\psib\equiv\T\psib$ and by the linearity of $\chr$ that
\[ \chr(\psib) = \frac{\chr(\psib_1)+\chr(\psib_2)}{2} = c \]
implying that an invariant flow of character $c$ exists.

The other degenerate case is when the $\mleb(F)=0$ condition is not satisfied by the IFS in Definition \ref{s03040001}. Again, the discussion cannot proceed. However, by the continuity of the invariant flow in the IFS parameters (Theorem \ref{s04030004}), we can approximate the invariant flow field of such a degenerate case arbitrarily closely with the invariant flows induced by approximating IFS parameters.
\section{The Model in Practice} \label{s04}

\subsection{The Evolution towards Invariance} \label{s0402}

As reasoned in Section \ref{s0302}, the intermittent evolution towards a fully developed turbulent flow field corresponds to the iteration of the weighted or probabilistic transfer operator towards the invariant flow $\psib$ characterized in Theorem \ref{s03050001}.

One of the advantages of this model over classical Chaotic Advection formalism, is that the intermittent or iterative steps need not occur at equal time intervals, and in fact the spacing in time can be arbitrary. An interesting question is whether the fully developed state can be achieved in finite time. The answer is, it certainly can. If the time spacing of each iteration decreases say geometrically, then the total time to reach the $L=\infty$ level iteration (corresponding to $\psib$) will be finite. This is clearly true for any convergent sequence of time spacing.

Another question which may arise when considering the previous sections, is how the global superposition of $n$ pushforwarded stream functions according to the transfer operator $\T$ relates to the local experimental picture of the splitting of eddies. Upon some contemplation, one may realize that the global and local viewpoints are the direct consequence of the associativity of transfer $\T(\T^L\psi_{c,p}) = \T^L(\T\psi_{c,p})$. Executing an iteration of $\T$ over the entire flow field $\T^L\psi_{c,p}$ corresponds to the local intermittent splitting of the initial eddy $\psi_{c,p}$ into $n$ eddies $\T\psi_{c,p}$ and therefore the splitting of all of its level $L$ iterates.

\subsection{The Inverse Problem} \label{s0403}

We now turn to translating the well-known inverse problem of IFS to fractal potential flows, in order to introduce a method for representing flow fields via this model in practice. The necessary Collage Theorem is detailed below, as well as the continuity of the invariant flow in the IFS parameters. Lastly, some numerical solution approaches are listed from the literature.

\begin{prb} \label{s04030001} \textup{(Inverse Problem - Applications)}
Given a finite number of intermittently stirred eddies, model their resulting interaction region as closely as possible.
\end{prb}

In terms of the introduced model, this vaguely stated problem becomes the following more precise formulation, based on articles \cite{bp00002} and further detailed in \cite{ba00066, bb00009}.

\begin{prb} \label{s04030002} \textup{(Inverse Problem - Theoretical)}
Given a target flow $\psi\in\PsXc$ and an $\ve>0$ accuracy, find an IFS of similarity contractions $\mcT=\{T_1,\ldots,T_n\}$ and corresponding weights $w_1,\ldots,w_n$ for some given $n\in\N$, such that for the induced invariant flow $\psib$ we have
\[ D(\psi,\psib)= \sup_{f\in\Lip}\ \frac{1}{\gc}\int_X f\ \lap(\psi-\psib)\ <\ve. \]
\end{prb}

The computation of $D$ can be carried out according to existing efficient methods for the Hutchinson metric \cite{ba00045, ba00064}, considering that $D(\psi_1,\psi_2)=d(\mu[\psi_1],\mu[\psi_2])$. As we have seen in Theorem \ref{s03040002}, the measure map $\mu$ is an isometric isomorphism between measures $\PX$ and flows $\PsXc$ (for some fixed $c$ character). Viewing each element of $\PX$ as a (potentially infinite) convex combination of Dirac deltas, the metric $D$ measures the difference between flows according to the power of their constituent sink singularities. So the inverse problem is about matching singularities. The rest of the flow field is only consequential, due to Poisson's Equation and the boundary condition called character (Theorem \ref{s03030003}). We now interpret a result by Barnsley et al. \cite{bp00002}.

\begin{thm} \label{s04030003} \textup{(Collage Theorem)}
Let $\psi\in\PsXc$ be a target flow, and suppose that there exists an IFS $\mcT=\{T_1,\ldots,T_n\}$ and weights $w_1,\ldots,w_n$ for some $n\in\N$ such that
\[ D(\psi,\T\psi) < (1-\vl_*)\ve\ \ \mathrm{with}\ \ \vl_*:=\max_{k\in\mcN} \vl_k \]
where $\T = w_1 T_1^* +\ldots + w_n T_n^*$. Then for its attractor $\psib\equiv \T\psib$ we have $D(\psi,\psib)<\ve$.
\end{thm}\begin{proof}
The theorem follows trivially from the triangle inequality for $D$.
\end{proof}

The above theorem allows us to design algorithms for resolving the inverse problem, which focus on improving or minimizing the distance $\T\mapsto D(\psi,\T\psi)$ over all transfer operators (i.e. varying IFS parameters and weights), towards the target flow $\psi$. Then the attractor $\psib$ of this approximate minimizer $\T$ will be ``close'' to $\psi$, meaning it solves the inverse problem up to the required $\ve$ accuracy. The Collage Theorem thus greatly simplifies this resolution, as originally discussed in \cite{bp00002}. The following is a straightforward translation of the result by Centore and Vrscay \cite{ba00047}.

\begin{thm} \label{s04030004} \textup{(Continuity of the Invariant Flow in Transfer Parameters)}\\
Let $\mcT_{X,c}$ be the set of all transfer operators $\T:\PsXc\rightarrow\PsXc$ with respect to any IFS and weights. Taking any two transfer operators $\T_{1,2}\in\mcT_{X,c}$ with contraction factors $\vl_{1,2}\in (0,1)$ and corresponding invariant flows $\psib_{1,2}\in\PsXc$ then
\[ D(\psib_1,\psib_2)\ \leq\ \frac{1}{1-\min(\vl_1,\vl_2)}\ \sup_{\psi\in\PsXc} D(\T_1(\psi),\T_2(\psi)). \]
\end{thm}

This theorem tells us that if the transfer operators $T_1$ and $T_2$ are close in the above supremum metric, then so will their attracting invariant flows $\psib_{1,2}$ be. This ensures the stability of the aforementioned process of minimization.

The solution approaches to the Inverse Problem come in many forms. They often rely on the fact that convergence in moments implies convergence in the Hutchinson metric, based on the idea of ``moment matching'' \cite{ba00017, ic00001, bu00008, ba00065}, oft-performed via gradient methods \cite{ic00001}, simulated annealing with memory \cite{ba00046}, and genetic algorithms \cite{ba00066}. Other methods include ``exact methods'' for solving the IFS inverse problem \cite{ba00012}, as well as the methods of Barnsley and Iterated Systems Inc., reported by Lu \cite{bb00011}. For a recent survey of the aforementioned techniques, see \cite{bb00009}.

\subsection{Experimental Verification} \label{s0404}

One may be inclined to test the physical correctness of the model experimentally, regardless of the fact that the previously detailed inverse problem alone supports its relevance to the practical modeling of turbulent flow fields. So in this section, we focus on isolating the elements of the model which must be critically investigated experimentally. Let us recall that Section \ref{s03} defined the chaotic mixing of $n$ intermittently stirred eddies as the iteration of the weighted transfer operator $\T = w_1 T_1^* +\ldots + w_n T_n^*$ over flows of a certain character $c\in\C$, so we must primarily focus on verifying the experimental validity of this iteration. The implicit assumptions of the model are as follows.
\begin{enumerate}
\item Fully Developed Turbulence may be generated by the intermittent interaction of $n\in\N$ eddies on a free surface.
\item In the local universe of the interacting eddies, the flow has an overall ``eddy character''.
\item The eddies are stirred intermittently by an external energy source along the logarithmic spiral trajectories of $T_k$ (meaning each eddy action can be represented by $T_k^*$).
\item At an iterative stir $L\in\N$ the transfer evolution occurs as follows.
\begin{enumerate}
\item The next stream value is determined by a discrete random variable $\hat{\psi}_{L+1}(z)$ at each $z\in\C$, with possible values $T_1^*\hat{\psi}_L(z),\ldots,T_n^*\hat{\psi}_L(z)$ and probabilities $w_1,\dots,w_n$.
\item Eddy $T_k^*$ affects the flow field $\psi$ according to the chaos game and Markov property
\[ \mP\left(\hat{\psi}_{L+1}(z)=T_k^*\psi_L(z)\ \mpbar\ \hat{\psi}_L=\psi_L\right)=w_k\ \ \ (z\in\C,\ k=1,\ldots,n,\ L\in\N) \]
\[ \mP\left(\hat{\psi}_{L+1}(z)=s_{L+1}\ \mpbar\ \hat{\psi}_L(z)=s_L,\ldots,\hat{\psi}_1(z)=s_1\right) = \]
\[ = \mP\left(\hat{\psi}_{L+1}(z)=s_{L+1}\ \mpbar\ \hat{\psi}_L(z)=s_L\right) \]
the latter expressing that the next state $\hat{\psi}_{L+1}(z)$ only depends on the current one.
\item The individual eddy actions merge collectively at each $z\in\C$ according to
\[ \psi_{L+1}(z):= \mE\left[\hat{\psi}_{L+1}(z)\ \mpbar\ \hat{\psi}_L=\psi_L\right] = \]
\[ = \sum_{k=1}^n T_k^*\psi_L(z)\ \mP\left(\hat{\psi}_{L+1}(z)=T_k^*\psi_L(z)\ \mpbar\ \hat{\psi}_L=\psi_L\right)=\T\psi_L(z). \]
\end{enumerate}
\end{enumerate}

Assumption 1. was argued in Section \ref{s0206}, while 2. seems reasonable, considering the scaling property of the Navier-Stokes Equations (if the velocity field $v(x,t)$ is a solution then so is $\frac{1}{\vl}v(\frac{x}{\vl}, \frac{t}{\vl^2})$), which implies that eddies should superpose as an eddy from a global vantage point - this is essentially ``character''. Assumption 3. means that each eddy receives its energy from a source external to their isolated local universe, either from the ``energy cascade'' or by direct stirring. That an intermittent eddy action can be represented by the pushforward transfer $T_k^*$ depends on whether 4. holds. Clearly 4. (a, c) follow directly from (b). So the experimental verification of the model reduces to 4. (b) which is an interpretation of Ott's Chaos Game \cite{ba00038} detailed in Section \ref{s020302}.

Restating 4. (b) succinctly by Theorem \ref{s03010001}: Is it true that at any locus $z\in\C$ on the free surface, the horizontal component $v_{L+1}$ of the next surface velocity field depends only on the current one $v_L$ via the following probabilities?
\[ \mP\left(\hat{v}_{L+1}(z)=\frac{1}{\vl_k}\me^{\vt_k\vi}\ v_L\left(T_k^{-1}(z)\right)\ \mpbar\ \hat{v}_L=v_L\right) = w_k\ \ \ (z\in\C,\ k=1,\ldots,n,\ L\in\N). \]
This property is only likely to hold on a small scale, where turbulent yet smooth stretching and folding of the fluid, transitions to non-differentiable intermittent eddy interactions due to the arising relevance of sweeping molecular effects at near-infinite Reynolds numbers.

\subsection{Visualization} \label{s0401}

\begin{figure}[H]
\centering
\includegraphics[width=400pt]{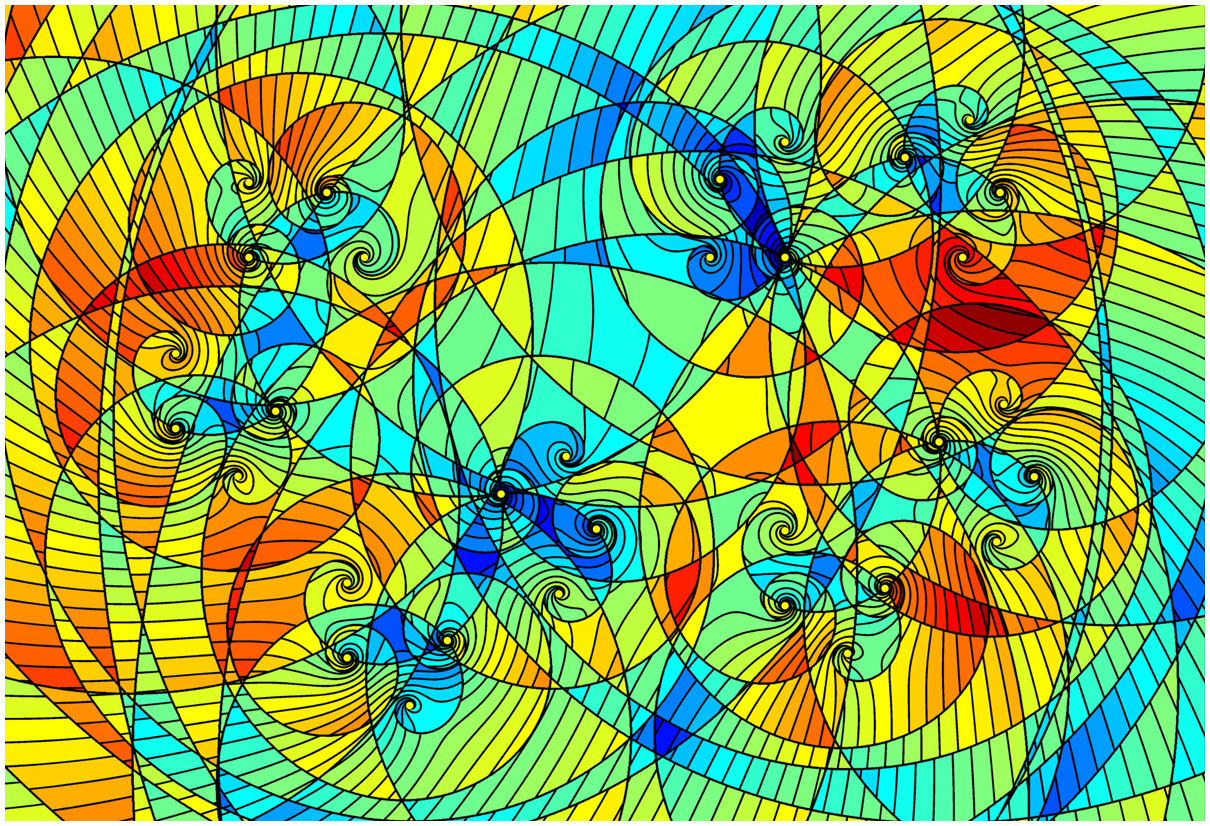}
\caption{Level curves of the stream function $\T^4\psi_0$ with the parameters given in Section \ref{s0401}, also indicating its logarithmic spiral branch cuts.}
\label{s04010001}
\end{figure}

\begin{figure}[H]
\centering
\vspace{0.15cm}
\includegraphics[width=400pt]{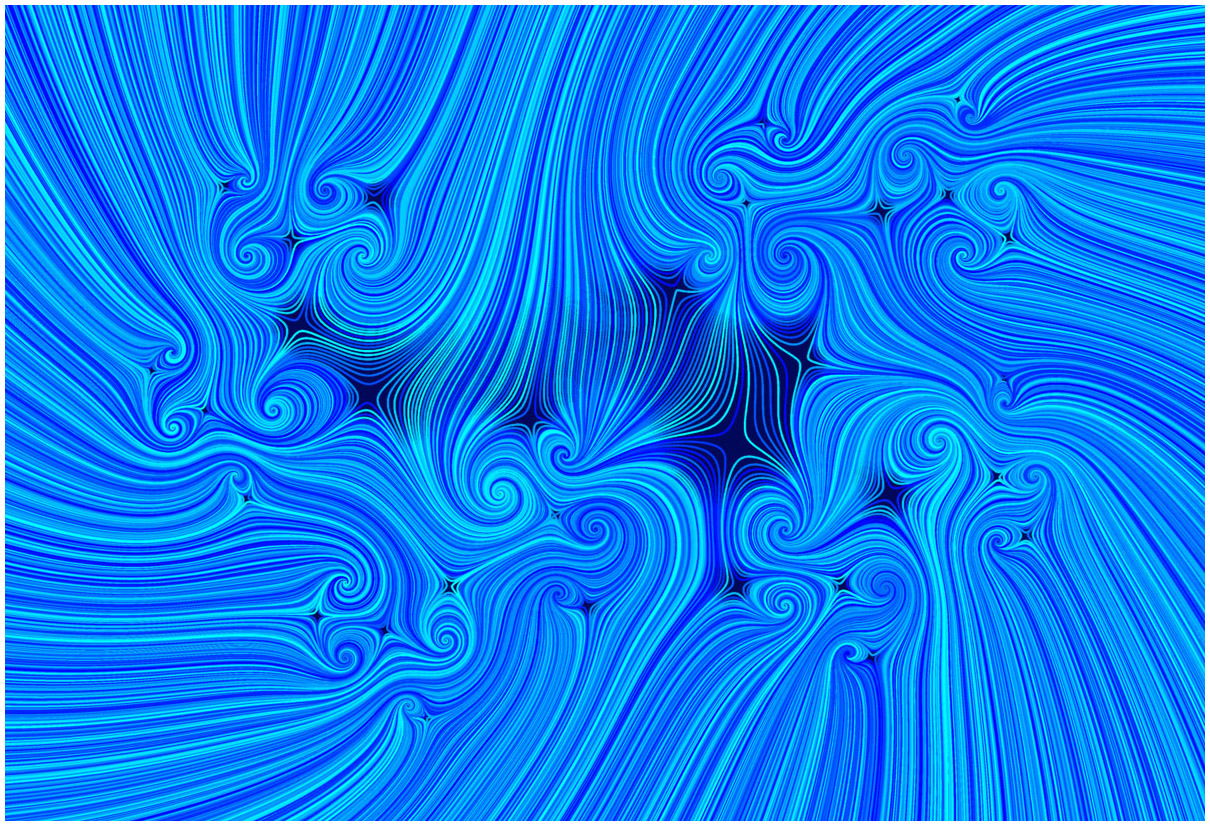}
\caption{The same streamlines solved from the velocity field.}
\label{s04010002}
\end{figure}

\begin{figure}[H]
\centering
\includegraphics[width=400pt]{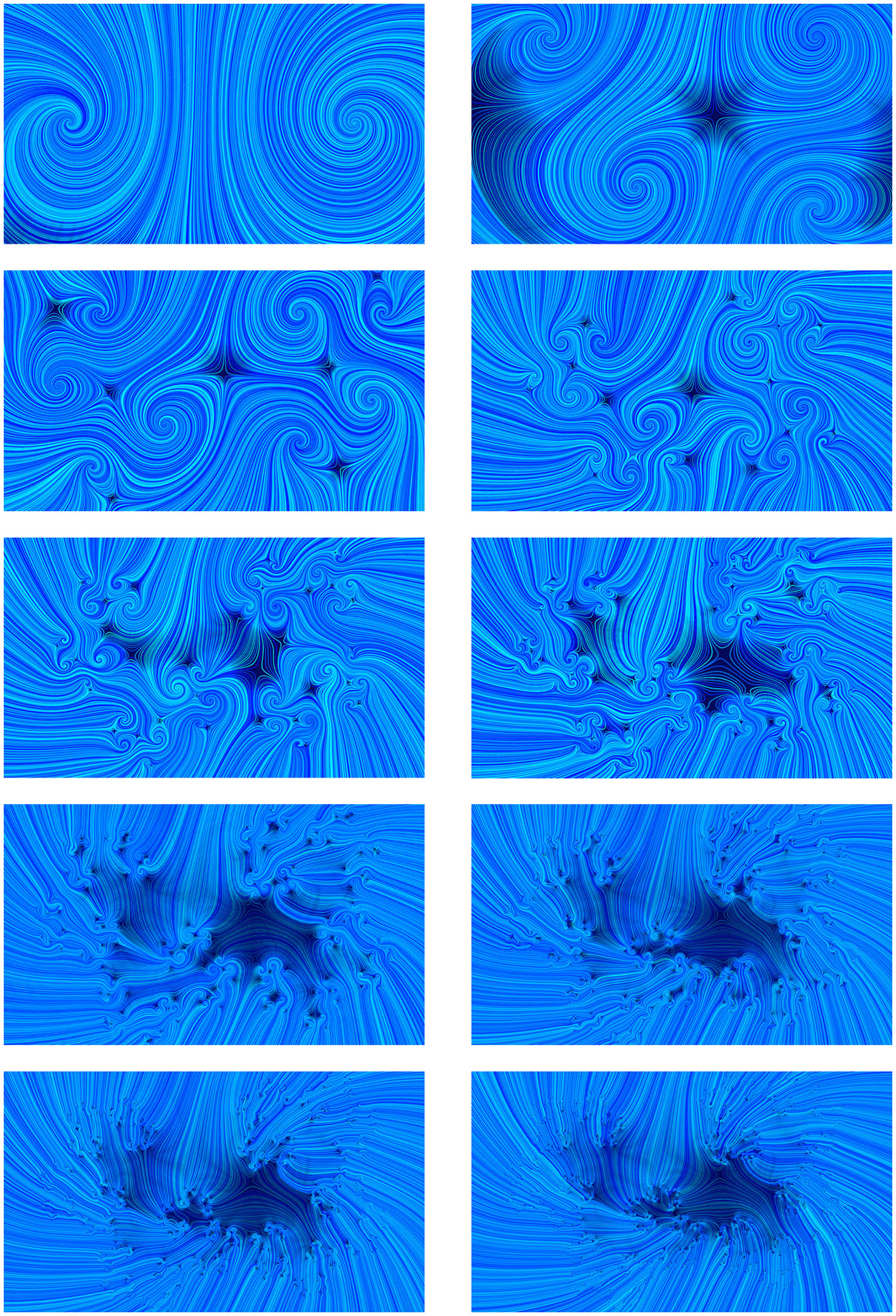}
\vspace{0.25cm}
\caption{Converging transfer iteration, plotted from $\psi_0$ to $\T^9\psi_0$.}
\label{s04010003}
\end{figure}

\begin{figure}[H]
\centering
\includegraphics[width=400pt]{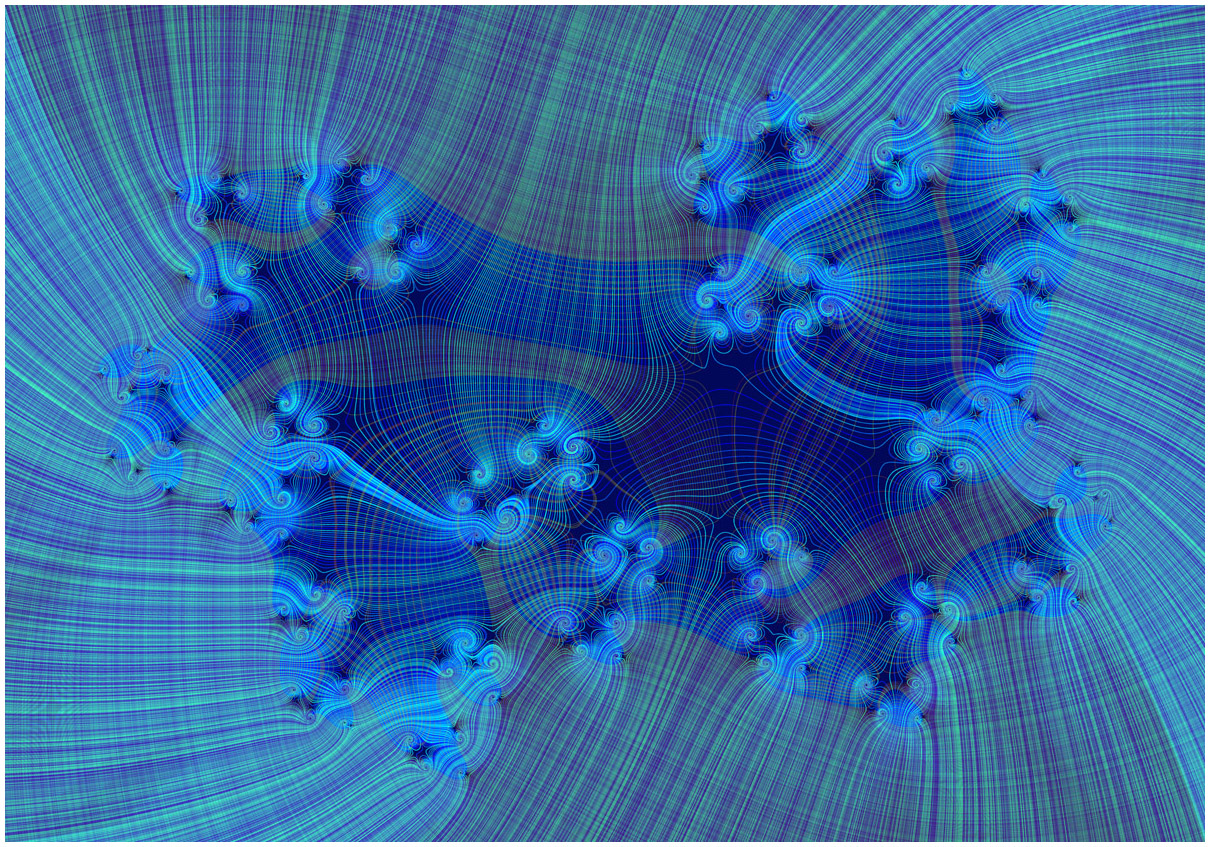}\\
\vspace{0.8cm}
\includegraphics[width=400pt]{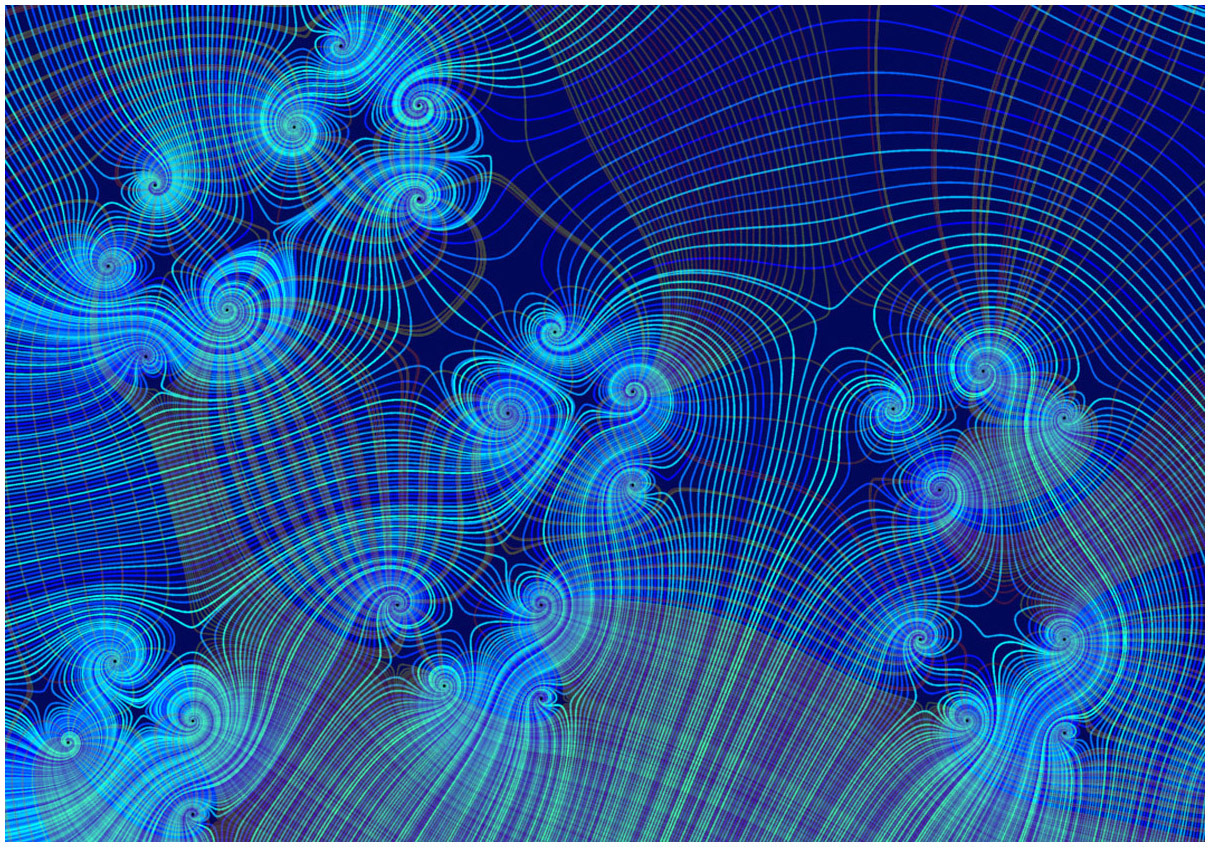}\\
\vspace{0.25cm}
\caption{The equipotential lines at iteration level $L=6$, and a detail image.}
\label{s04010004}
\end{figure}

When considering the transfer operator over stream functions, one encounters a proliferation of branch cuts as the transfer iteration progresses. This can make the visualization of streamlines - the level curves of the stream function - quite difficult. Note that stream functions in $\bPsXc$ are only required to be almost everywhere smooth, and the transfer $\T$ produces a finite Hutchinson union of new branch cuts at each iteration, which pose no issue theoretically, despite the visual and algorithmic mess they create in practice. This is illustrated by Fig. \ref{s04010001} with transfer parameters
\[ p_1=0,\ p_2=1,\ \cf_1=0.65 \me^{-\frac{2\pi}{6}\vi},\ \cf_2=0.65\me^{\frac{2\pi}{4}\vi},\ w_{1,2}=0.5. \]
Therefore in order to visualize the flow field, one seems to have no option but to resort to an iteration over the velocity field. $\psi_0$ can be chosen arbitrarily, as apparent from Theorem \ref{s03050001}, so we choose $\psi_0:=\sum_k w_k T_k^*\psi_{c_k,p_k}$ as discussed at the end of Section \ref{s0304}, where $c_k=\Log\ \cf_k$ (choosing $C=2\pi$ in Section \ref{s0301}) for an IFS of Section \ref{s0204}. The evolution in the velocity field progresses according to Theorem \ref{s03020001}. So we have the following recursive iteration for the velocity field, over the flow space $(\bPsXc, D)$ with preserved character $c=\sum_k w_k\Log\ \cf_k$ over the compact region $X\supset\langle T_1,\ldots,T_n\rangle$.
\[ v_0(z)= v[\psi_0] = \sum_{k=1}^n w_k(\Log\ \cf_k)\frac{z-p_k}{|z-p_k|^2},\ \ v_{L+1}(z)= v[\T\psi_L](z)= \sum_{k=1}^n w_k \frac{\cf_k}{|\cf_k|^2}\ v_L(T_k^{-1}(z)). \]
Once at a high enough iteration level $L\in\N$, one may execute a streamline solver algorithm, resulting in a preferable image as in Fig. \ref{s04010002}. The transfer iteration is visualized in Fig. \ref{s04010003}. Note the apparent convergence to an attractor flow field, as shown in Section \ref{s03}. In order to visualize the level curves of the corresponding potential function - the equipotential lines - first note the correspondence in character $\chr(\tilde{\psi})= -\vi\ \chr(\psi)$ by Definition \ref{s03030001}. If one does not wish to deal with the arising sources in the conjugate to the above $\psi_0$ (as in Fig. \ref{s04010004}), we may choose simply instead $\psi_0:=\psi_{c,p}$ and thus $\tilde{\psi}_0=\psi_{-\vi c,p}$ with the primary fixed point $p$ and the above $c$, as seen in Section \ref{s0305}. Then in order to arrive at the potential field at level $L$, we execute the same velocity recursion on $\tilde{v}_0(z):=v[\tilde{\psi}_0]$ as above on $v_0(z)= v[\psi_0]$. The iterative formula remains the same because of Theorems \ref{s03010001} and \ref{s03020001}, implying the following with $\tilde{v}_L:=v[\tilde{\psi}_L],\ L\in\{0\}\cup\N$
\[ \tilde{v}_0(z)= v[\tilde{\psi}_{0}](z)= -\vi c\ \frac{z-p}{|z-p|^2},\ \ \tilde{v}_{L+1}= v[\tilde{\psi}_{L+1}]= v[\widetilde{\T\psi_L}]= v[\T\tilde{\psi}_L]= \sum_{k=1}^n w_k\frac{\cf_k}{|\cf_k|^2}\ T_k^*\tilde{v}_L. \]
Depending on the sign of $\mIm\ c$, it may be more convenient from the viewpoint of a streamline solver, to take the negative of $\tilde{v}_0$ so that $\mRe\ \chr(\tilde{v}_0)<0$.

\section{Concluding Remarks} \label{s05}



To suggest future research, one may find the conjecture stated in Section \ref{s030602} worth examining, as well as the inspirational remarks on entropy in Section \ref{s0305}. Furthermore, the isometric isomorphism introduced in Section \ref{s0304} between flows and probability measures, may be exploited to translate former results on measures to the language of flows, while examining the physical implications. Considering Section \ref{s030601}, the study of Fractal Potential Flows reduces essentially to the geometry of planar IFS fractals with similarity contractions, so their investigation appears to bear fundamental relevance. Section \ref{s0403} detailing the inverse problem may prove its worth in modeling flow fields of a certain character, regardless of whether experiments verify or disprove the model, in light of Section \ref{s0404}.

The author is grateful to Stephen Wiggins, Julio M. Ottino, and Tamás Tél for their preliminary reviews of the first draft, to Morten Br\o ns for a discussion on the aforementioned conjecture, and to John C. Sommerer for clarifications of his work with Edward Ott. The mathematical model of Section \ref{s03} was a part of the author's doctoral dissertation \cite{ph00004}. The paper is dedicated to the late Benoît B. Mandelbrot and Hassan Aref.

\bibliographystyle{abbrv}
\bibliography{mybib}

\begin{thebibliography}{10}

\bibitem{ba00034}
H.~Aref.
\newblock Stirring by chaotic advection.
\newblock {\em Journal of Fluid Mechanics}, 143:1--21, 1984.

\bibitem{ba00035}
H.~Aref, S.~Jones, S.~Mofina, and I.~Zawadzki.
\newblock Vortices, kinematics, and chaos.
\newblock {\em Physica D}, 37:423--440, 1989.

\bibitem{bb00002}
M.~F. Barnsley.
\newblock {\em Fractals Everywhere}.
\newblock Academic Press, Boston, second edition, 1993.

\bibitem{ba00017}
M.~F. Barnsley and S.~Demko.
\newblock Iterated function systems and the global construction of fractals.
\newblock {\em Proceedings of the Royal Society}, A399, 1985.

\bibitem{bp00002}
M.~F. Barnsley, V.~Ervin, D.~Hardin, and J.~Lancaster.
\newblock Solution of an inverse problem for fractals and other sets.
\newblock In {\em Proceedings of the National Academy of Sciences of the United
  States of America}, volume~83, pages 1975--1977. National Academy of
  Sciences, 1986.

\bibitem{ba00092}
G.~Batchelor and A.~Townsend.
\newblock The nature of turbulent motion at large wave-numbers.
\newblock {\em Proceedings of the Royal Society of London, Series A},
  199(1057):238--255, 1949.

\bibitem{bb00013}
R.~B. Bird, W.~E. Stewart, and E.~N. Lightfoot.
\newblock {\em Transport phenomena}.
\newblock John Wiley \& Sons, second edition, 2007.

\bibitem{bp00007}
M.-E. Brachet and S.~A. Orszag.
\newblock Secondary instability of free shear flows.
\newblock In {\em Turbulence and Chaotic Phenomena in Fluids}, volume~1, pages
  27--29, 1984.

\bibitem{ba00045}
J.~Brandt, C.~Cabrelli, and U.~Molter.
\newblock An algorithm for the computation of the {H}utchinson distance.
\newblock {\em Information processing letters}, 40(2):113--117, 1991.

\bibitem{bu00008}
C.~Cabrelli, U.~Molter, and E.~Vrscay.
\newblock Moment matching for the approximation of measures using iterated
  function systems.
\newblock 1992.

\bibitem{ba00047}
P.~Centore and E.~R. Vrscay.
\newblock Continuity of attractors and invariant measures for iterated function
  systems.
\newblock {\em Canad. Math. Bull}, 37(3):315--329, 1994.

\bibitem{bb00014}
H.~Chanson.
\newblock {\em Applied hydrodynamics: {A}n introduction to ideal and real fluid
  flows}.
\newblock CRC Press, 2009.

\bibitem{ba00012}
A.~Deliu, J.~Geronimo, and R.~Shonkwiler.
\newblock On the inverse fractal problem for two-dimensional attractors.
\newblock {\em Philosophical Transactions of the Royal Society of London,
  Series A}, 355(1726):1017--1062, 1997.

\bibitem{bp00001}
R.~L. Devaney.
\newblock Subshifts of finite type in linked twist mappings.
\newblock In {\em Proceedings of the American Mathematical Society}, volume~71,
  pages 334--338, 1978.

\bibitem{ba00064}
V.~Drakopoulos and N.~P. Nikolaou.
\newblock Efficient computation of the {H}utchinson metric between digitized
  images.
\newblock {\em Image Processing, IEEE Transactions on}, 13(12):1581--1588,
  2004.

\bibitem{ba00032}
R.~Ecke.
\newblock The turbulence problem, an experimentalist's perspective.
\newblock {\em Los Alamos Science}, (29):124--140, 2005.

\bibitem{ba00065}
B.~Forte and E.~R. Vrscay.
\newblock Solving the inverse problem for measures using iterated function
  systems: {A} new approach.
\newblock {\em Advances in applied probability}, pages 800--820, 1995.

\bibitem{ba00087}
U.~Frisch.
\newblock Fully developed turbulence and intermittency.
\newblock {\em Annals of the New York Academy of Sciences}, 357(1):359--367,
  1980.

\bibitem{bb00016}
U.~Frisch.
\newblock {\em Turbulence: The Legacy of A. N. Kolmogorov}.
\newblock Cambridge University Press, 1996.

\bibitem{ba00077}
U.~Frisch, P.-L. Sulem, and M.~Nelkin.
\newblock A simple dynamical model of intermittent fully developed turbulence.
\newblock {\em Journal of Fluid Mechanics}, 87(4):719--736, 1978.

\bibitem{ba00098}
G.~Froyland, K.~Padberg, M.~H. England, and A.~M. Treguier.
\newblock Detection of coherent oceanic structures via transfer operators.
\newblock {\em Physical Review Letters}, 98(22):224503, 2007.

\bibitem{ba00007}
J.~E. Hutchinson.
\newblock Fractals and self similarity.
\newblock {\em Indiana University Mathematics Journal}, 30:713--747, 1981.

\bibitem{ba00033}
D.~Jou.
\newblock Intermittent turbulence: A short introduction.
\newblock {\em Scientia Marina}, 61:57--62, 1997.

\bibitem{ba00041}
L.~Kantorovich and G.~Rubinstein.
\newblock On the space of completely additive functions.
\newblock {\em Ser. Mat. Mekh. i Astron.}, 13(7):52--59, 1958.

\bibitem{ba00049}
J.~Kigami.
\newblock A harmonic calculus on the {S}ierpinski spaces.
\newblock {\em Japan Journal of Applied Mathematics}, 6(2):259--290, 1989.

\bibitem{ba00050}
J.~Kigami.
\newblock Harmonic calculus on pcf self-similar sets.
\newblock {\em Transactions of the American Mathematical Society},
  335(2):721--755, 1993.

\bibitem{bb00010}
J.~Kigami.
\newblock {\em Analysis on fractals}, volume 143.
\newblock Cambridge University Press, 2001.

\bibitem{bp00004}
A.~N. Kolmogorov.
\newblock The local structure of turbulence in incompressible viscous fluid for
  very large {R}eynolds numbers.
\newblock In {\em Dokl. Akad. Nauk SSSR}, volume~30, pages 299--303, 1941a.

\bibitem{bp00005}
A.~N. Kolmogorov.
\newblock On the degeneration of isotropic turbulence in an incompressible
  viscous fluid.
\newblock In {\em Dokl. Akad. Nauk SSSR}, volume~31, pages 538--541, 1941b.

\bibitem{bp00006}
A.~N. Kolmogorov.
\newblock Dissipation of energy in locally isotropic turbulence.
\newblock In {\em Dokl. Akad. Nauk SSSR}, volume~32, pages 16--18, 1941c.

\bibitem{ba00082}
A.~N. Kolmogorov.
\newblock A refinement of previous hypotheses concerning the local structure of
  turbulence in a viscous incompressible fluid at high reynolds number.
\newblock {\em Journal of Fluid Mechanics}, 13(1):82--85, 1962.

\bibitem{ic00007}
R.~H. Kraichnan.
\newblock In S.~Rice, K.~Freed, and J.~Light, editors, {\em Statistical
  mechanics; new concepts, new problems, new applications}, page 201.
  University of Chicago Press, 1972.

\bibitem{ba00081}
R.~H. Kraichnan.
\newblock On {K}olmogorov's inertial-range theories.
\newblock {\em Journal of Fluid Mechanics}, 62(2):305--330, 1974.

\bibitem{ba00036}
G.~Károlyi and T.~Tél.
\newblock Chaotic tracer scattering and fractal basin boundaries in a blinking
  vortex-sink system.
\newblock {\em Physics Reports}, 290:125--147, 1997.

\bibitem{bb00009}
H.~Kunze, D.~L. Torre, F.~Mendivil, and E.~Vrscay.
\newblock {\em Fractal-based methods in analysis}.
\newblock Springer, 2012.

\bibitem{ba00097}
A.~Y.-S. Kuo and S.~Corrsin.
\newblock Experiments on internal intermittency and fine-structure distribution
  functions in fully turbulent fluid.
\newblock {\em Journal of Fluid Mechanics}, 50(02):285--319, 1971.

\bibitem{ba00089}
J.~Leray.
\newblock Sur le mouvement d'un liquide visqueux emplissant l'espace.
\newblock {\em Acta Mathematica}, 63(1):193--248, 1934.

\bibitem{bb00011}
N.~Lu.
\newblock {\em Fractal imaging}.
\newblock Morgan Kaufmann Publishers Inc., 1997.

\bibitem{bc00008}
B.~B. Mandelbrot.
\newblock On intermittent free turbulence.
\newblock Turbulence of Fluids and Plasmas, New York, April 1968. Polytechnic
  Institute of Brooklyn, Interscience.

\bibitem{ba00080}
B.~B. Mandelbrot.
\newblock Intermittent turbulence in self-similar cascades: {D}ivergence of
  high moments and dimension of the carrier.
\newblock {\em Journal of Fluid Mechanics}, 62(2):331--358, 1974.

\bibitem{ic00006}
B.~B. Mandelbrot.
\newblock Turbulence and {N}avier-{S}tokes equation.
\newblock In R.~Temam, editor, {\em Lecture Notes in Mathematics}, volume 565,
  page 121. Springer, 1976.

\bibitem{ba00046}
G.~Mantica and A.~Sloan.
\newblock Chaotic optimization and the construction of fractals: {S}olution of
  an inverse problem.
\newblock {\em Complex Systems}, 3:37--62, 1989.

\bibitem{bc00009}
R.~J. McCann and N.~Guillen.
\newblock Five lectures on optimal transportation: Geometry, regularity and
  applications.
\newblock In {\em Analysis and Geometry of Metric Measure Spaces: Lecture Notes
  of the 50th Séminaire de Mathématiques Supérieures (SMS)}, Montréal, 2011.

\bibitem{ba00083}
W.~D. McComb.
\newblock The physics of fluid turbulence.
\newblock {\em Chemical Physics}, 1, 1990.

\bibitem{ba00078}
C.~Meneveau and J.~Katz.
\newblock Scale-invariance and turbulence models for large-eddy simulation.
\newblock {\em Annual Review of Fluid Mechanics}, 32(1):1--32, 2000.

\bibitem{bb00018}
A.~S. Monin, A.~M. Yaglom, and J.~L. Lumley.
\newblock {\em Statistical Fluid Mechanics: Mechanics of Turbulence}, volume~2.
\newblock MIT Press, 1975.

\bibitem{ba00079}
E.~Novikov and R.~Stewart.
\newblock The intermittency of turbulence and the spectrum of energy
  dissipation fluctuations.
\newblock {\em Izv. Geophys. Ser}, 3:408--413, 1964.

\bibitem{ba00088}
S.~A. Orszag and A.~T. Patera.
\newblock Subcritical transition to turbulence in plane channel flows.
\newblock {\em Physical review letters}, 45:989--993, 1980.

\bibitem{bb00008}
J.~M. Ottino.
\newblock {\em The kinematics of mixing: stretching, chaos, and transport},
  volume~3.
\newblock Cambridge University Press, 1989.

\bibitem{ba00076}
J.~M. Ottino.
\newblock Mixing, chaotic advection, and turbulence.
\newblock {\em Annual Review of Fluid Mechanics}, 22(1):207--254, 1990.

\bibitem{ba00085}
O.~Reynolds.
\newblock On the dynamical theory of incompressible viscous fluids and the
  determination of the criterion.
\newblock {\em Philosophical Transactions of the Royal Society of London,
  Series A}, pages 123--164, 1895.

\bibitem{bb00015}
L.~F. Richardson.
\newblock {\em Weather prediction by numerical process}.
\newblock Cambridge University Press, 1922.

\bibitem{ba00094}
L.~F. Richardson.
\newblock Atmospheric diffusion shown on a distance-neighbour graph.
\newblock {\em Proceedings of the Royal Society of London, Series A},
  110(756):709--737, 1926.

\bibitem{bb00017}
J.~P. Richter.
\newblock {\em The notebooks of Leonardo da Vinci}, volume~2.
\newblock Courier Dover Publications, 1970.

\bibitem{bb00019}
W.~Rudin.
\newblock {\em Principles of mathematical analysis}, volume~3.
\newblock McGraw-Hill New York, 1964.

\bibitem{ba00090}
D.~Ruelle.
\newblock Measures describing a turbulent flow.
\newblock {\em Annals of the New York Academy of Sciences}, 357(1):1--9, 1980.

\bibitem{ba00091}
D.~Ruelle and F.~Takens.
\newblock On the nature of turbulence.
\newblock {\em Communications in mathematical physics}, 20(3):167--192, 1971.

\bibitem{ba00086}
V.~Scheffer.
\newblock The {N}avier-{S}tokes equations on a bounded domain.
\newblock {\em Communications in Mathematical Physics}, 73(1):1--42, 1980.

\bibitem{ba00095}
E.~D. Siggia.
\newblock Origin of intermittency in fully developed turbulence.
\newblock {\em Physical Review A}, 15(4):1730, 1977.

\bibitem{ba00096}
E.~D. Siggia.
\newblock Model of intermittency in three-dimensional turbulence.
\newblock {\em Physical Review A}, 17(3):1166, 1978.

\bibitem{ba00040}
J.~C. Sommerer.
\newblock The end of classical determinism.
\newblock {\em Johns Hopkins APL Technical Digest}, 16(4):333--347, 1995.

\bibitem{ba00039}
J.~C. Sommerer and E.~Ott.
\newblock Particles floating on a moving fluid: A dynamically comprehensible
  physical fractal.
\newblock {\em Science, New Series}, 259(5093):335--339, 1993.

\bibitem{ba00037}
J.~C. Sommerer, E.~Ott, and T.~Tél.
\newblock Modeling two-dimensional fluid flows with chaos theory.
\newblock {\em Johns Hopkins APL Technical Digest}, 18(2):193--203, 1997.

\bibitem{ba00084}
K.~R. Sreenivasan.
\newblock Fractals and multifractals in fluid turbulence.
\newblock {\em Annual Review of Fluid Mechanics}, 23(1):539--604, 1991.

\bibitem{ba00051}
R.~S. Strichartz.
\newblock Analysis on fractals.
\newblock {\em Notices AMS}, 46(10):1199--1208, 1999.

\bibitem{ba00093}
G.~I. Taylor.
\newblock The spectrum of turbulence.
\newblock {\em Proceedings of the Royal Society of London, Series A},
  164(919):476--490, 1938.

\bibitem{ph00004}
J.~Vass.
\newblock {\em On the Geometry of IFS Fractals and its Applications}.
\newblock PhD thesis, University of Waterloo, 2013.

\bibitem{ba00099}
J.~Vass.
\newblock On intersecting {IFS} fractals with lines.
\newblock {\em Fractals}, 2014.
\newblock Accepted.

\bibitem{bb00007}
C.~Villani.
\newblock {\em Optimal Transport: Old and New}, volume 338.
\newblock Springer, 2008.

\bibitem{ic00005}
J.~von Neumann.
\newblock Recent theories of turbulence.
\newblock In A.~H. Taub, editor, {\em Collected Works (1949-1963)}, volume~6,
  page 437. Pergamon Press, Oxford, 1963.
\newblock Original publication 1949.

\bibitem{ba00066}
E.~R. Vrscay.
\newblock Iterated function systems: {T}heory, applications and the inverse
  problem.
\newblock {\em Fractal Geometry and Analysis}, pages 405--468, 1991.

\bibitem{ic00001}
E.~R. Vrscay and C.~J. Roehrig.
\newblock Iterated function systems and the inverse problem of fractal
  construction using moments.
\newblock In {\em Computers and Mathematics}, pages 250--259. Springer, 1989.

\bibitem{ba00043}
S.~Wiggins.
\newblock Chaos in the dynamics generated by sequences of maps, with
  applications to chaotic advection in flows with aperiodic time dependence.
\newblock {\em Zeitschrift f{\"u}r Angewandte Mathematik und Physik (ZAMP)},
  50(4):585--616, 1999.

\bibitem{ba00044}
S.~Wiggins and J.~M. Ottino.
\newblock Foundations of chaotic mixing.
\newblock {\em Philosophical Transactions of the Royal Society of London,
  Series A}, 362(1818):937--970, 2004.

\bibitem{ba00038}
L.~Yu, E.~Ott, and Q.~Chen.
\newblock Fractal distribution of floaters on a fluid surface and the
  transition to chaos for random maps.
\newblock {\em Physica D}, 53:102--124, 1991.

\end{thebibliography}

\end{document}